\documentclass[a4paper,9pt,twocolumn]{extarticle} 
\usepackage[super,sort&compress,comma]{natbib} 
\usepackage[version=3]{mhchem} 
\usepackage[left=1.5cm, right=1.5cm, top=1.785cm, bottom=2.0cm]{geometry}
\usepackage{balance} 
\usepackage{graphicx} 
\usepackage{lastpage}
\usepackage[format=plain,justification=justified,singlelinecheck=false,font={stretch=1.125,small,sf},labelfont=bf,labelsep=space]{caption}
\usepackage{float} 
\usepackage[english]{babel}
\addto{\captionsenglish}{%
  
}
\usepackage{array} 
\usepackage{charter} 
\usepackage[T1]{fontenc} 
\usepackage{xcolor} 
\usepackage{setspace} 
\usepackage[compact]{titlesec}
\usepackage{hyperref} 
\usepackage{amsmath,amssymb} 
\usepackage[normalem]{ulem} 

\definecolor{cream}{RGB}{222,217,201}

\newcommand{\mycirc}{^{\circ}}

\begin{document}


\makeatletter
\renewcommand\LARGE{\@setfontsize\LARGE{15pt}{17}}
\renewcommand\Large{\@setfontsize\Large{12pt}{14}}
\renewcommand\large{\@setfontsize\large{10pt}{12}}
\renewcommand\footnotesize{\@setfontsize\footnotesize{7pt}{10}}
\makeatother

\renewcommand{\thefootnote}{\fnsymbol{footnote}}
\renewcommand\footnoterule{\vspace*{1pt}%
\color{cream}\hrule width 3.5in height 0.4pt \color{black}\vspace*{5pt}} 
\setcounter{secnumdepth}{5}

\makeatletter 
\renewcommand\@biblabel[1]{#1}            
\renewcommand\@makefntext[1]%
{\noindent\makebox[0pt][r]{\@thefnmark\,}#1}
\makeatother 
\renewcommand{\figurename}{\small{Fig.}~}
\setstretch{1.125} 
\setlength{\skip\footins}{0.8cm}
\setlength{\footnotesep}{0.25cm}
\setlength{\jot}{10pt}
\titlespacing*{\section}{0pt}{4pt}{4pt}
\titlespacing*{\subsection}{0pt}{15pt}{1pt}

\setlength{\arrayrulewidth}{1pt}
\setlength{\columnsep}{6.5mm}
\setlength\bibsep{1pt}

\makeatletter 
\newlength{\figrulesep} 
\setlength{\figrulesep}{0.5\textfloatsep} 

\newcommand{\topfigrule}{\vspace*{-1pt}%
\noindent{\color{cream}\rule[-\figrulesep]{\columnwidth}{1.5pt}} }

\newcommand{\botfigrule}{\vspace*{-2pt}%
\noindent{\color{cream}\rule[\figrulesep]{\columnwidth}{1.5pt}} }

\newcommand{\dblfigrule}{\vspace*{-1pt}%
\noindent{\color{cream}\rule[-\figrulesep]{\textwidth}{1.5pt}} }

\makeatother

\twocolumn[
\begin{@twocolumnfalse}

\vspace{1em}
\sffamily

\noindent\LARGE{\textbf{
Sorting of binary active-passive mixtures in designed microchannels
}} \\

\noindent\large{
Horacio Serna\textit{$^{a,b}$}, 
C. Miguel Barriuso G.\textit{$^{a,b}$}, 
Ignacio Pagonabarraga\textit{$^{c,d}$},  
Marco Polin\textit{$^e$}\textsuperscript{*} 
and Chantal Valeriani\textit{$^{a,b}$}\textsuperscript{*}
}
\\

\noindent\normalsize{Mixtures of active and passive particles are ubiquitous at the microscale. Many essential microbial processes involve interactions with dead or immotile cells or passive crowders. When passive objects are immersed in active baths, their transport properties are enhanced and can be tuned by controlling active agents' spatial and orientational distribution. Active-passive mixtures provide a platform to explore fundamental questions about the emergent behaviour of passive objects under simultaneous thermal and active noise and a foundation for technological applications in cargo delivery and bioremediation. In this work, we use computational simulations to study an active-passive mixture confined in microchannels designed with funnel-like obstacles that selectively allow the passage of passive particles. Active particles follow overdamped Langevin translational dynamics and run-and-tumble rotational dynamics. We find that adjusting the tumbling rate of active agents and the microchannel geometry leads to a maximum enhancement of the transport properties of the passive particles (diffusion coefficient and advective velocity) that correlates with the highest mixture sorting efficiency and the shortest response time.} \\
\vspace{0.5em}
\noindent\textsuperscript{*}Corresponding authors: \texttt{mpolin@imedea.uib-csic.es}, \texttt{cvaleriani@ucm.es}
\end{@twocolumnfalse} \vspace{0.6cm}

]


\renewcommand*\rmdefault{bch}\normalfont\upshape
\rmfamily
\section*{}
\vspace{-1cm}



\footnotetext{\textit{$^{a}$~Departamento de Estructura de la Materia, Física Térmica y Electrónica, Universidad Complutense de Madrid, 28040 Madrid, Spain}}
\footnotetext{\textit{$^{b}$ GISC - Grupo Interdisciplinar de Sistemas Complejos 28040 Madrid, Spain}}
\footnotetext{\textit{$^{c}$ Departament de F\'isica de la Matèria  Condesada, Facultat de F\'isica - Universitat de Barcelona, Carrer de Mart\'i i Franquès, 1, 11, 08028 Barcelona, Spain}}
\footnotetext{ \textit{$^ {d}$ Universitat de
  Barcelona Institute of Complex Systems (UBICS), Universitat de Barcelona, 08028 Barcelona, Spain}}
\footnotetext{ \textit{$^ {e}$Instituto Mediterráneo de Estudios Avanzados (IMEDEA), CSIC-UIB. Miquel Marquès 21, 07190, Esporles, Balearic Islands, Spain}}



\section{Introduction}


The interaction between active and passive agents is ubiquitous and determines critical processes for microorganisms. Parasites moving in crowded environments to spread infection \cite{heddergott2012trypanosome}, motile microorganisms swimming in waterbodies' complex environments to feed and facilitate trophic fluxes \cite{montagnes2008selective}, and the swimming mechanism modulation of \textit{Euglena gracilis} in crowded media \cite{noselli2019swimming} are some examples. 

From a fundamental point of view, mixtures of active and passive particles constitute {a novel class of out-of-equilibrium thermodynamic systems, where the fluctuations 
shaking
the passive species have statistical properties that can be externally tuned at the microscale.}
{From an experimental perspective, active baths can be easily realized with} suspensions of motile microorganisms like bacteria \cite{wu2000particle,valeriani2011colloids,jepson2013enhanced,patteson2016particle} or microalgae \cite{jeanneret2016entrainment,leptos2009dynamics,kurtuldu2011enhancement}. Depending on the passive colloids' size, the self-propulsion speed of the microorganisms, and the concentration of cells in the active baths, 
the passive colloids' diffusion coefficient might {be orders of magnitude larger than the thermal one}
\cite{wu2000particle,patteson2016particle,jeanneret2016entrainment}. 
{Recently,}
synthetic active-passive mixtures have been studied in the context of the effects of crowders and external fields on active colloids' motion, showing that the complex interplay of hydrodynamics and chemical fields \citealp{hauke2020clustering,sturmer2019chemotaxis,madden2022hydrodynamically}, as well as collisions with passive agents \cite{singh2022interaction}, result in modulations of translational and rotational dynamics of  active colloids.

{Despite recent progress in probing} the behaviour of active-passive mixtures {from a
theoretical
\cite{wittkowski2017nonequilibrium,takatori2015theory}, numerical
\cite{ye2020active,dhar2024active,stenhammar2015activity,zhang2022density,semwal2024dynamics,maggi2017memory,shea2022passive,rodriguez2020phase,dolai2018phase,wang2020phase}, 
and experimental
\cite{wu2000particle,patteson2016particle,valeriani2011colloids,leptos2009dynamics,angelani2011effective,jepson2013enhanced,kurtuldu2011enhancement,jeanneret2016entrainment,seyforth2022nonequilibrium}
perspective,}
understanding their phase behaviour in bulk and under confinement and the mechanism behind the enhancement of the transport properties of  passive colloids, is still very much an open challenge
\cite{bechinger2016active,gompper20202020,gompper20252025}.

The  transport properties' enhancement of passive colloids can be modified by reorganizing the active bath so that the spatial and orientational distribution of active agents are non-homogeneous. Different methods may achieve this reorganization. The first method consists in modifying the shape of passive agents. Experimental and simulation studies have shown that passive elongated agents (polymers or colloidal chains) embedded in active baths induce a non-homogeneous local packing fraction of active particles close to them, resulting, in turn, in localized distributions of the passive agents within the bath and enhanced anisotropic diffusion \cite{aporvari2020anisotropic,mousavi2021active}. Designing the passive objects in the shape of spheres\cite{mallory2014curvature}, rods\cite{mallory2014curvature,ni2015}, chevrons \cite{kaiser2014transport,mallory2014curvature,wang2020different} or cogwheels \cite{di2010bacterial,sokolov2010swimming} induces the accumulation of  active agents in the concave regions of the objects, leading to a net translation and rotation, respectively. The net transport induced by the  shape might be exploited to develop engines powered by active agents \cite{pietzonka2019autonomous,fodor2021active}. 
The second method imposes external stimuli on  active agents, such as  light or spatial gradients of physico-chemical properties.
For example, light has been used to control the assembly of active colloids \cite{schmidt2019light}, tune the motion of Active Brownian Particles (ABP) \cite{buttinoni2012active}, and control the flow patterns of microalgae suspensions\cite{arrieta2019light,garcia2013light,yang2021controlling}. 

Finally, the third method is confinement, a simple way to generate a space-dependent distribution of active agents, affecting fluctuations and transport of passive particles. This is the method we explore in this article. Recently, it was experimentally demonstrated that spatial confinement leads to a non-uniform steady-state distribution of microalgae that accumulates at boundaries and induces de-mixing of passive particles \cite{williams2022confinement}. Studies have demonstrated that chevron- or funnel-like obstacles can effectively sort, trap, and separate active particles. For instance, the carnivorous plant Genlisea utilizes its rhizophylls' funnel-shaped structures to rectify bacterial swimming, directing them toward its digestive vesicle\cite{martin2025carnivorous}. At the same time, microfabricated funnel walls have been shown to concentrate swimming bacteria by guiding their motion through the funnels 
\cite{galajda2007wall}. Trapping of self-propelled particles with alignment interactions was observed between channels composed of funnel walls\cite{martinez2020trapping}, as well as  trapping of self-propelled rods by chevron-like obstacles\cite{kaiser2012capture,kumar2019trapping}.

By using the methods described above, one can leverage the enhancement of the transport properties and the eventual induced net transport of passive particles to develop technological applications in micro cargo-delivery and bioremediation \cite{yasa2018microalga,martinez2015magnetic,yang2020cargo,fu2022microscopic}. Most studies on active-passive mixtures consider passive colloids in bacterial baths with smaller active agents than passive ones. Much less studied are baths of eukaryotic microorganisms that can be selected to be larger than the passive agents. Inspired by the experiments with confined mixtures of the biflagellate microalgae \textit{C. reinhardtii} and colloidal beads \cite{jeanneret2016entrainment,williams2022confinement}, we consider an active-passive mixture whose active agents are larger than the passive species.
The size difference  might be exploited in sorting or separation processes at the microscale powered by active matter. 

\begin{figure*}[t!]
 \centering
 \includegraphics[width=14cm]{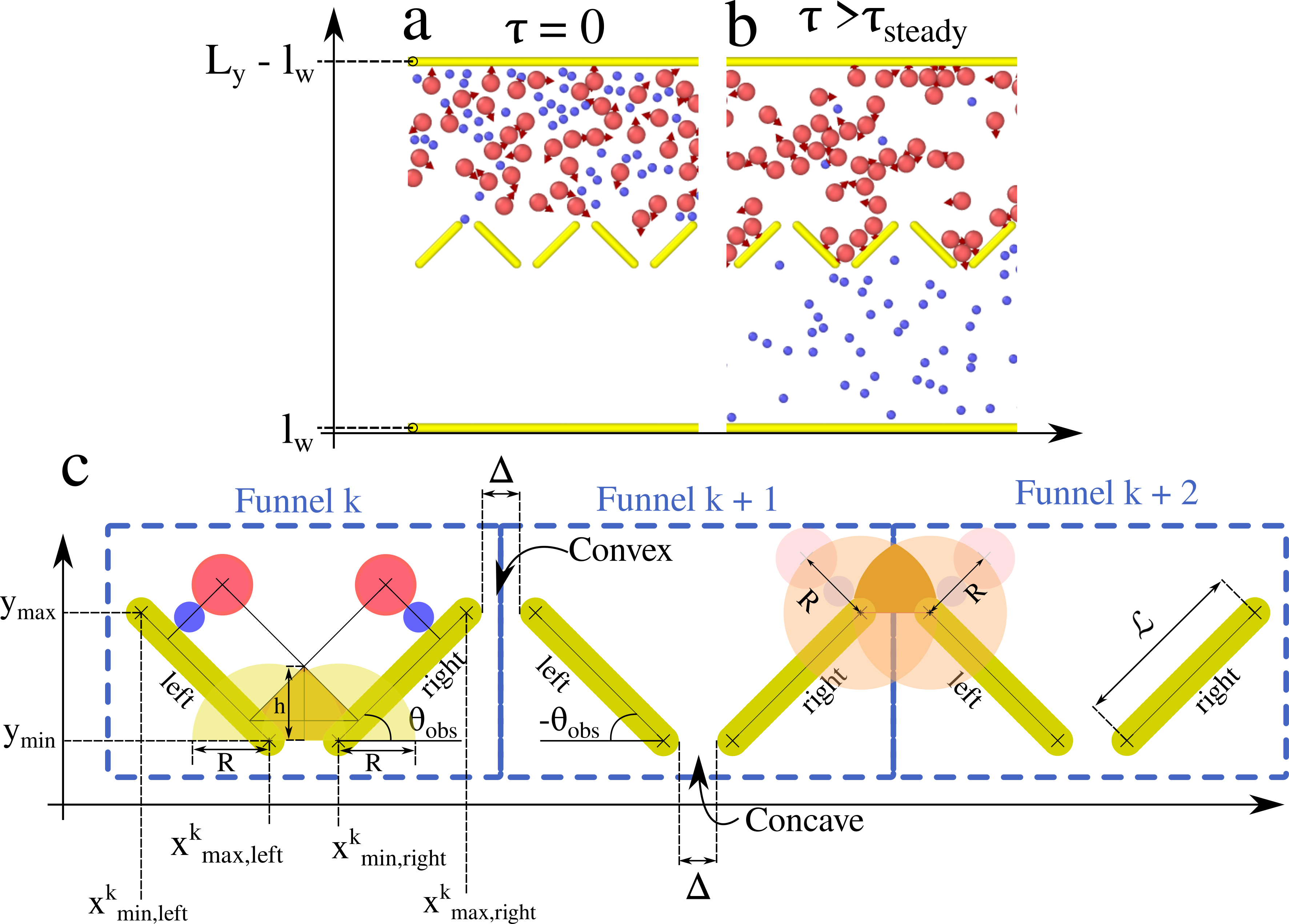}
 \caption{\textit{The system and geometry of the microchannel}. Pink particles are active, and their orientation vector is represented in dark red. Blue particles are passive, and yellow particles are frozen and constituents of the channel's walls. The funnel-like obstacles in the central region of the channel have gaps that only allow the passage of passive particles. The active-passive binary mixture is confined in the $y$-axis by placing walls at $y = l_w$ and $y = L_y - l_w$. Periodic boundary conditions are applied along the $x$-axis. {\bf{(a)}} The particles are initially placed in random positions in the top compartment. {\bf{(b)}} At the steady state, the sorting of the mixture is achieved; the concentration of passive particles in the bottom compartment is much higher than that of the top compartment.{\bf{(c)}} The gap between obstacles is $\Delta = 1.25\sigma_{22}$ and the angle between the obstacles and the $x$-axis is $\theta_{obs}$. The length of the obstacles is $\mathcal{L} = 3\sigma_{11}$. The reference for the concave and convex regions is the top compartment of the microchannel. The blue dashed-line rectangles represent the funnels composed of two barriers labeled as "left" and "right". For the concave gaps, the clogging region for $\theta_{obs}\leq30^{\circ}$ is that consisting of the intersection of the yellow semi-circles of radius, $R = \sigma_{33}/2 +\sigma_{22} + \sigma_{11}/2$, in the funnel $k$, and for $\theta_{obs}>30\mycirc$ the clogging area is that consisting of the area below the straight lines parallel to the barriers at a distance $R$ from the center of the barrier and depicted in dark orange in funnel $k$. For the convex gaps, the clogging region is defined as the intersection of 2 circles placed at the center of the extreme wall particles with radius, $R$. All the graphics of trajectories of simulations presented in this article were generated in part using the visualization software Ovito \cite{stukowski2009visualization}.}
 \label{fgr:System}
\end{figure*}

It has been shown that the boundaries of either the passive species -when sufficiently large- \cite{mallory2014curvature} or of the confining space \cite{ostapenko2018curvature}, can influence the statistical properties of the active bath. 
Thus, it stands to reason that, in turn, they will also regulate the transport of passive agents. 
Here we explore 
the latter case by confining the active-passive mixture in a microchannel with funnel-like obstacles in its central part (Fig.~\ref{fgr:System}).
By systematically changing the curvature of the boundary and the diffusivity of the active species, our results reveal the existence of boundary shape parameters that are optimal for the expulsion of passive particles. This phenomenon is then rationalised in terms of the ratio between the obstacles' size and the run distance of the active agents.

The article is organized in the following way. In Section \ref{Numerical}, the model and the analysis tools are described. Next, in Section \ref{Results}, the results are presented and discussed. Finally, in Section \ref{Conclusions}, we present the conclusions and   perspectives for future work.

\section{Numerical details}

\label{Numerical}

\subsection{Model}

The system under study consists of a two-dimensional binary mixture of active (type 1) and passive (type 2) particles, confined in a designed microchannel composed of frozen particles (type 3) (See Figure\ref{fgr:System}). 
Active particles (in pink in Figure \ref{fgr:System}) are simulated as run-and-tumble particles, whereas passive particles  (in blue in Figure \ref{fgr:System}) as Brownian particles.

Run-and-tumble is implemented as follows. Active particles
 obey a Langevin dynamics,
\begin{equation}
    m_i\frac{d^2 \textbf{r}_i}{dt^2} = F_a\textbf{n}_i -  \boldsymbol{\nabla}_i U -\gamma\,\textbf{v}_i + \sqrt{2\gamma m_i k_BT}\,\textbf{W}_i
    \label{e:langevin}
\end{equation}
where $m_i$, $\textbf{r}_i$, $\textbf{n}_i$, $\textbf{v}_i$, are the mass,  position vector,  unit orientation vector, and  velocity vector of particle $i$, respectively. $F_a$ is the magnitude of the active force that acts along the orientation of the active particles, and $\gamma$ is the friction coefficient of the implicit solvent in which particles are embedded. For active particles, we set $F_a = 10$, while $F_a = 0$ corresponds to the case of passive particles (blue particles in Figure \ref{fgr:System}). $k_B$ is the Boltzmann constant and $T$ is the temperature of the underlying thermal bath. We set $k_B T = 1$ in all studied cases. This generates a thermal force $\textbf{W}_i=(W_i^x,W_i^y)$ acting on the $i$th particle, given by the standard Gaussian white noise with zero mean and correlation $\langle W_i^x(t)W_j^y(t') \rangle = \delta_{ij}\delta_{xy}\delta(t-t')$.
%
We choose particles to interact repulsively via a WCA\cite{weeks1971role} potential,
\begin{equation}
    U(r) =
    \begin{cases} 
       4\epsilon_{\alpha\beta}\left[ \left ( \frac{\sigma_{\alpha\beta}}{r} \right)^{12} -\left (\frac{\sigma_{\alpha\beta}}{r} \right)^6 \right] + \epsilon_{\alpha\beta}, & r< 2^{1/6}\sigma_{\alpha\beta}, \\
        0 & r \geq 2^{1/6}\sigma_{\alpha\beta}.
    \end{cases}
    \label{e:WCA}
\end{equation}
Here, the indexes $\alpha$ and $\beta$ run over the three particle types: active, passive, and wall particles. We set $\sigma_{11} = 2$, $\sigma_{22} = 1$, and $\epsilon_{11} = \epsilon_{22} = 1$. The Lorentz-Berthelot mixing rules\cite{lorentz1881ueber,berthelot1898melange}, $\sigma_{\alpha\beta} = \left(\sigma_{\alpha\alpha} + \sigma_{\beta\beta}\right)/2$ and $\epsilon_{\alpha\beta} = \sqrt{\epsilon_{\alpha\alpha}\epsilon_{\beta\beta}}$, are used when particles of different type interact. The particles constituting the microchannel's walls (type 3, in yellow in Figure \ref{fgr:System}) do not interact with each other but interact with the active (type 1) and passive (type 2) particles via the same WCA potential and following the mixing rules with $\epsilon_{33} = 1$ and $\sigma_{33} = 1$. 
Throughout the article, all  quantities will be expressed in reduced units using $k_BT$, $\sigma_{22}$ and $m_2$ as energy, distance, and mass units, respectively.

Since the active and passive particles have different sizes but are embedded in the same solvent, we can write the following relation as a consequence of Stokes' Law, $\gamma_1/\gamma_2 = \sigma_1/\sigma_2 = 2$. Thus, we set $\gamma_1 = 10$ and $\gamma_2 = 5$.  With these values of the friction coefficients, the system is close to the overdamped regime:  even though  inertial effects are present at short times,  they are small enough not to be the most important contribution to the system's dynamics (\textbf{see Fig. S1}). Besides, solving the full Langevin's equation \ref{e:langevin} is a more general approach that applies to active-passive mixtures in a broader context where inertial effects are present \cite{sprenger2023dynamics,deblais2018boundaries},  
even though the latter is not the scope of the presented work.

Regarding the rotational dynamics,    active particles follow a run-and-tumble motion, a common swimming phenotype observed in bacteria such as \textit{E. coli}\cite{berg1985physics,kurzthaler2024characterization}, \textit{B. subtilis}\cite{ordal1985complementation,turner2016visualizing}, and \textit{Enterobacter Sp. SM3}\cite{johnson2024run}, but also, to a first approximation, in eukaryotic microorganisms such as \textit{C. reinhardtii}\cite{polin2009chlamydomonas}.
The following discrete stochastic process determines the frequency of tumbling events
\begin{equation}
\theta_i(t + dt) = 
    \begin{cases} 
       \theta_i(t), & \alpha < \zeta {(t)} \\
        \theta_i(t) + \Phi{(t)}  & \alpha > \zeta {(t)}
    \end{cases} ,
    \label{e:run-and-tumble}
\end{equation}
where $\theta_i$ is the angle formed between the orientation vector $\textbf{n}_i$ of particle $i$ and the positive $x$-axis. $\alpha = t_{trial}/\tau_r$, is the tumbling rate,  $t_{trial}$ being the time between trials, and $\tau_r$  the mean time between tumbling events or reorientation time. This model uses $t_{trial} = dt$, i.e. tumbling trials are performed at every step. $\zeta \in [0,1]$ is a uniformly distributed random number and $\Phi \in [0,2\pi]$ is a uniformly distributed random angle.

The system is composed of a total number of $N = 3000$ particles with $N_1 = \chi_{active}N$ and $N_2 = N - N_1$, where $\chi_{active}$ is the fraction of active particles in the mixture, here fixed at $0.50$.
In Figure \ref{fgr:System}, we present the system and the geometry of the obstacles considered in the article.  The snapshots are portions of the system to facilitate visualization.
As shown in the figure, the simulation box has a rectangular shape with height   $L_y = 40\sigma_{22}$ along the $y$-axis and width along the $x$-axis chosen such as to maintain the packing fraction of the mixture, $\phi = \frac{N_1\pi \sigma_{11}^2 + N_2\pi \sigma_{22}^2}{4L_xL_y} = 0.15$.
The system is initialised with all  Types 1 and 2 particles  randomly placed  in the top compartment of the microchannel (see Fig.\ref{fgr:System}a).
We impose periodic boundary conditions along the $x$-axis, while along the $y$-axis the system is confined by two parallel walls at $y = l_w$ and $y = L_y - l_w$. Here $l_w = 0.12$ is the separation between particles forming the walls (see Figure \ref{fgr:System}). 

In the central region, the microchannel has funnel-like obstacles featuring gaps that only allow the passage of the passive particles. 
Under these conditions, the microchannel is elongated in the $x$-direction and short in the $y$-axis, thus favoring the particles' encounters with the obstacles. 
Figure~\ref{fgr:System}c illustrates the geometry of the obstacles: their length is set to $\mathcal{L} = 3\sigma_{11}$, and they form an angle $\theta_{obs}$ with respect to the $x$-axis, which will be varied in the range $[0\mycirc,75\mycirc]$ in steps of $15\mycirc$.
The width $\Delta$ of the gaps between the obstacles is kept constant in all simulations, at $\Delta = 1.25\,\sigma_{22}$. This size is chosen to allow the flow of only one passive particle at a time. 
The motion of active particles induces a flow of passive particles towards the bottom compartment such that, once  steady state is reached, the concentration of passive particles is much higher at the bottom than at the top of the channel (see Fig.~\ref{fgr:System}b).

Simulations are performed using the open-source Molecular Dynamics  package LAMMPS \cite{plimpton1995fast} in which we have implemented the run-and-tumble motion described in equation \ref{e:run-and-tumble}. The Langevin's thermostat and the NVE integrator are used with $dt = 0.001$.  Simulations are run for $20000$ simulation time units, being $\tau=\sqrt{\frac{m_2\sigma_{22}^2}{k_BT}}$. Averages of all computed quantities are taken at steady state. Videos of the system for different values of $\theta_{obs}$ are presented in the $\textbf{Supplementary Video}$.

\subsection{Analysis tools}


To quantify the passive particles sorted from the mixture 
and thus the separation efficiency, we compute the average fraction of passive particles in the bottom compartment at steady state: 
%
\begin{equation}
    \langle\mathcal{F}_{bottom}\rangle = \frac{1}{n_{conf}N_2}\sum^{n_{conf}}_{\alpha=1} n_{2}(\alpha\tau).
    \label{e:TrappedFraction}
\end{equation}
Here $n_2(\alpha\tau)$ is the number of passive (type 2) particles whose position $\mathbf{x}_2(\alpha\tau)$ at time $\alpha\tau$ is below the obstacles. We sample $n_{conf} = 12000$ configurations to estimate the steady state value of $\langle\mathcal{F}_{bottom}\rangle$, which are separated by $\tau = 1$ simulation time units (1000 simulation time steps). 

One of the factors  determining the transition of passive particles across a gap is the likelihood of it being closed by an active particle. We will therefore define a gap as closed whenever an active particle is in a position that, geometrically, would not allow the passage of passive particles between the top and bottom sides of the chamber. The criterion depends on the intersection of exclusion zones defined around each one of the slanted obstacles forming the gaps. An exclusion zone is simply defined as the set of points that are closer than $\sigma_{11}/2 + \sigma_{22}$ to the surface of the obstacle. Whenever the centre of an active particle is within the exclusion region, passive particles cannot fit within the gap between the surface of the active particle and the obstacle. As a consequence, whenever the centre of an active particle is within the area at the intersection of the two exclusion zones of adjacent obstacles, or clogging area $A_c$, it is impossible for a passive particle to cross the corresponding gap (see Fig.~\ref{fgr:System}c). That gap is then considered as clogged. Examples $A_c$'s for different values of $\theta_{obs}$ can be found in \textbf{Fig.~S2}.  
Notice that, for large angles, the clogging regions for concave gaps are large enough to host more than one active particle. Since only one active particle is required to clog a gap, we consider 
the occupied gaps, not  the active particles inside the clogging regions. With this information, we can calculate the fraction of open concave and convex gaps, $\mathcal{P}_{concave}$ and $\mathcal{P}_{convex}$, as well as the total fraction of open gaps, $\mathcal{P}_{total} = \frac{\mathcal{P}_{concave} + \mathcal{P}_{convex}}{2}$.



\section{Results}
\label{Results}

\begin{figure}[t]
 \centering
 \includegraphics[width=7.0cm]{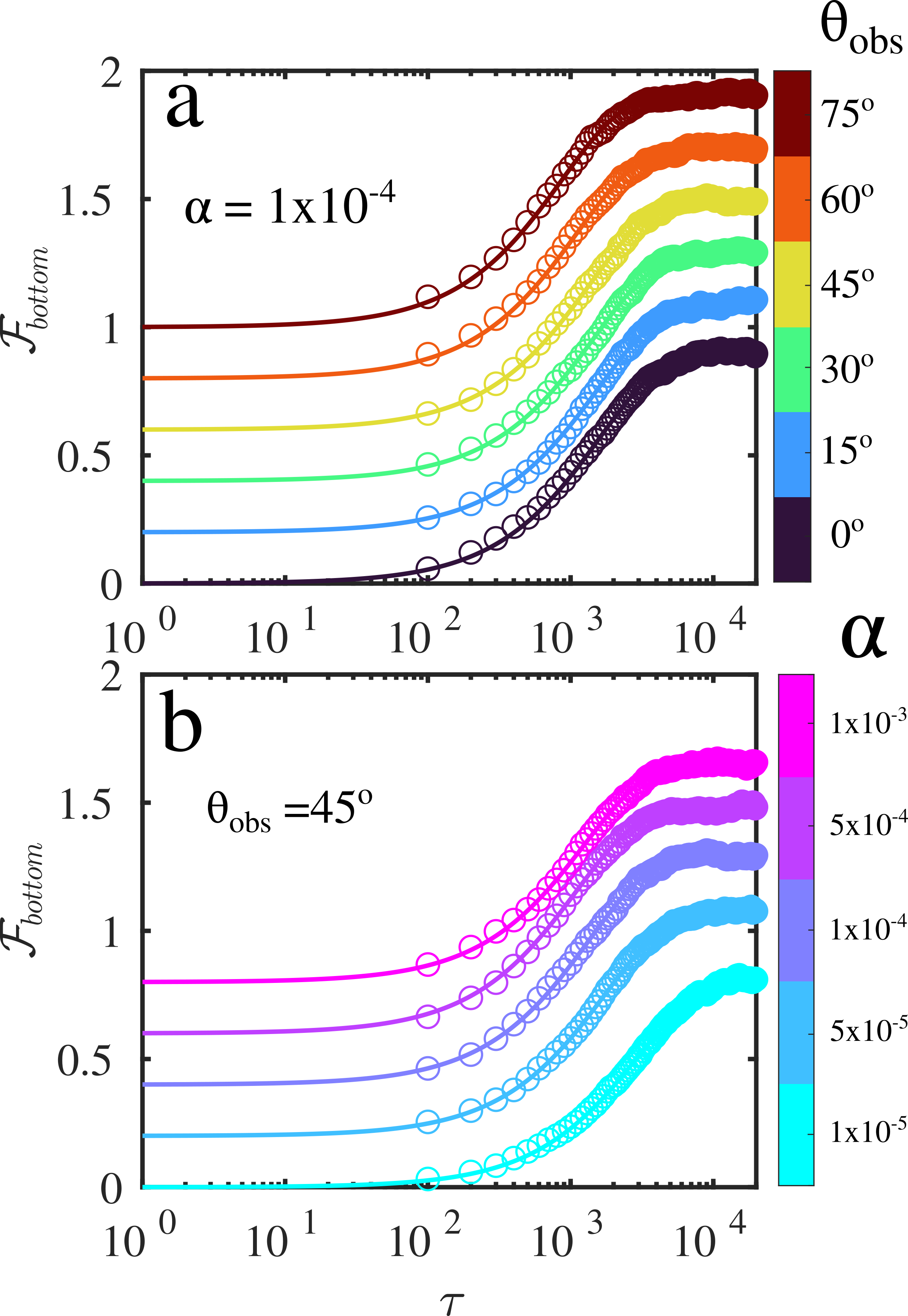}
 \caption{\textit{The sorting process follows first-order dynamics}. The simulation data is represented with circles, and the fits to the first-order dynamics are expressed in equation \ref{e:sol_N_bottom} in solid lines. The curves are represented in a semi-log scale in the $x$ axis and have a positive shift of $0.20$ in the $y$-axis for clarity.  \textbf{(a)} The fraction of passive particles in the bottom compartment, $\mathcal{F}_{bottom}$ as a function of time. The active particles' tumbling rate, $\alpha = 1\times10^{-4}$, is kept constant, whereas the angle of the obstacles $\theta_{obs}$ is varied. \textbf{(b)} $\mathcal{F}_{bottom}$ as a function of time when the angle of the obstacles, $\theta_{obs} = 45\mycirc$ and $\alpha$ is changing.}
 \label{fgr:FirstOrderDynamics}
\end{figure}

Figure \ref{fgr:FirstOrderDynamics} presents the time evolution of the fraction of passive particles in the bottom compartment of the microchannel, $\mathcal{F}_{bottom}$, calculated using eq. \ref{e:TrappedFraction}, keeping constant either the tumbling rate (panel a) or the angle between obstacles (panel b). 

\begin{figure*}[h]
 \centering
 \includegraphics[width=18.0cm]{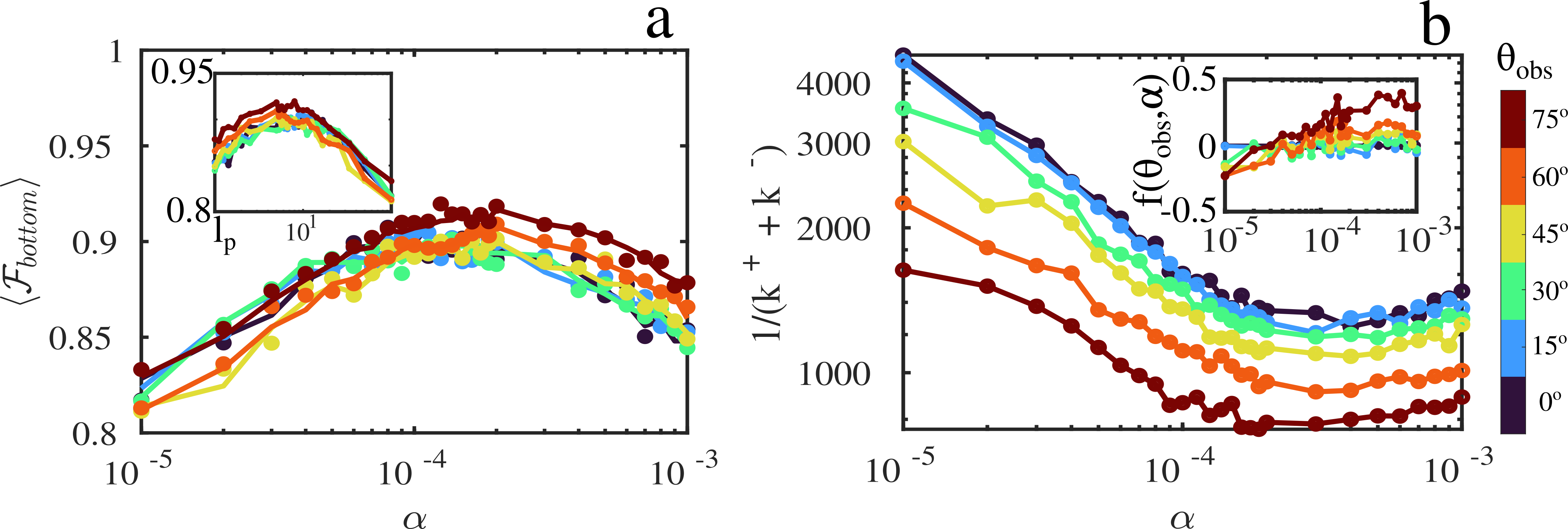}
 \caption{\textit{The dependence of the dynamic parameters on the activity and geometry.} \textbf{(a)} the average fraction of passive particles in the bottom compartment at steady state, $\langle \mathcal{F}_{bottom} \rangle$ and \textbf{(b)} the time constant of the sorting process as functions of the tumbling rate of the active particles, $\alpha$, and the angle of the obstacles $\theta_{obs}$. In \textbf{(a)} solid lines are predictions of the model in eq. \ref{e:ODE_first_order} given by $\mathcal{F}_{bottom}= \frac{k^+\mathcal{F}_{top}}{k^-}$ and in \textbf{(b)} are guides for the eye. The inset in \textbf{(a)} is $\langle\mathcal{F}_{bottom}\rangle$ as a function of the persistence length, $l_p = \frac{F_a dt}{\alpha\gamma_1}$. The inset in \textbf{(b)} is $f(\theta_{obs},\alpha)$ as presented in \ref{eq:taud} . Figure \textbf{(a)} is presented in a semi-log scale in the $x$-axis and figure \textbf{(b)} in a log-log scale.}
 \label{fgr:KineticConstants}
\end{figure*}

The evolution of $\mathcal{F}_{bottom}$ follows a simple first-order kinetic resulting from the mass balance of the passive particles at the top and bottom compartments of the microchannel:
%
\begin{equation}
\begin{cases}
\frac{d\mathcal{F}_{top}}{dt} &= -k^+\mathcal{F}_{top} + k^-\mathcal{F}_{bottom}\\
\frac{d\mathcal{F}_{bottom}}{dt} &=k^+\mathcal{F}_{top} - k^-\mathcal{F}_{bottom}
\end{cases},
\label{e:ODE_first_order}
\end{equation}
where $\mathcal{F}_{top/bottom}$ is the fraction of passive particles at the top/bottom compartment ($\mathcal{F}_{top} + \mathcal{F}_{bottom} = 1$);  $k^-$ is the kinetic constant determining the flow of passive particles from bottom to top and $k^+$ from top to bottom. 
%
%
%
The solution of the system of equations \ref{e:ODE_first_order} reads
\begin{equation}
    \mathcal{F}_{top}(t)=\frac{k^- + k^+e^{-t/\tau_d}}{k^+ + k^-},
    \label{e:sol_N_top}
\end{equation}
\begin{equation}
    \mathcal{F}_{bottom}(t)= \frac{k^+ \left( 1 - e^{-t/\tau_d}\right)}{k^+ + k^-},
    \label{e:sol_N_bottom}
\end{equation}
where the time constant is given by $\tau_d = 1/(k^+ + k^-)$. At steady state
$\mathcal{F}_{bottom}= \mathcal{F}_{top}k^+/k^-$. 

In Fig.~\ref{fgr:FirstOrderDynamics}, the circles correspond to data from the simulations, and the solid lines are fits to eq.~\ref{e:sol_N_bottom},  in excellent agreement with the simulation data. Figure~\ref{fgr:FirstOrderDynamics}a shows the variation of the dynamic response upon changes in the angle of the obstacles, $\theta_{obs}$, while the tumbling rate is kept constant at $\alpha = 1\times10^{-4}$. Conversely, in Fig.~\ref{fgr:FirstOrderDynamics}b the angle $\theta_{obs} = 45\mycirc$ is kept constant while $\alpha$ is varied. Clearly, for both sweeps, the fits capture the system's dynamics quite well. 

Although the simple kinetic model does not retain information on neither   the geometry of microchannels nor the active bath, 
it is enough to reconstruct the dynamics of the passive particles flowing through the funnels. The contribution of both  activity and geometry is encapsulated within the kinetic constants $k^+$ and $k^-$. Note that due to the asymmetric nature of the system, $k^+$ determines the flow of passive particles due to the enhancement of their transport properties when embedded in an active bath, while $k^-$ determines the flow of passive particles only due to thermal motion. Still, as we will see, $k^-$ is dependent on the behaviour of active particles, through the changes in opening and closing of the gaps. The results of the kinetic constants $k^+$ and $k^-$ are reported in \textbf{Fig. S3}. We observe that $k^+$ features a maximum that shifts to the left as $\theta_{obs}$ increases, from $\alpha \approx 4\times10^{-4}$ for $\theta_{obs} = 0\mycirc$, to $\alpha \approx 2\times10^{-4}$ for $\theta_{obs} = 75\mycirc$ (see \textbf{Fig. S3a}). The latter is also the global maximum of $k^+$, meaning that the flow of passive particles to the bottom compartment is maximized for those values of $\alpha$ and $\theta_{obs}$. Conversely, as seen in \textbf{Fig S3b}, $k^-$ increases monotonically with both $\alpha$ and $\theta_{obs}$.

\begin{figure*}[h!]
 \centering
 \includegraphics[width=18.0cm]{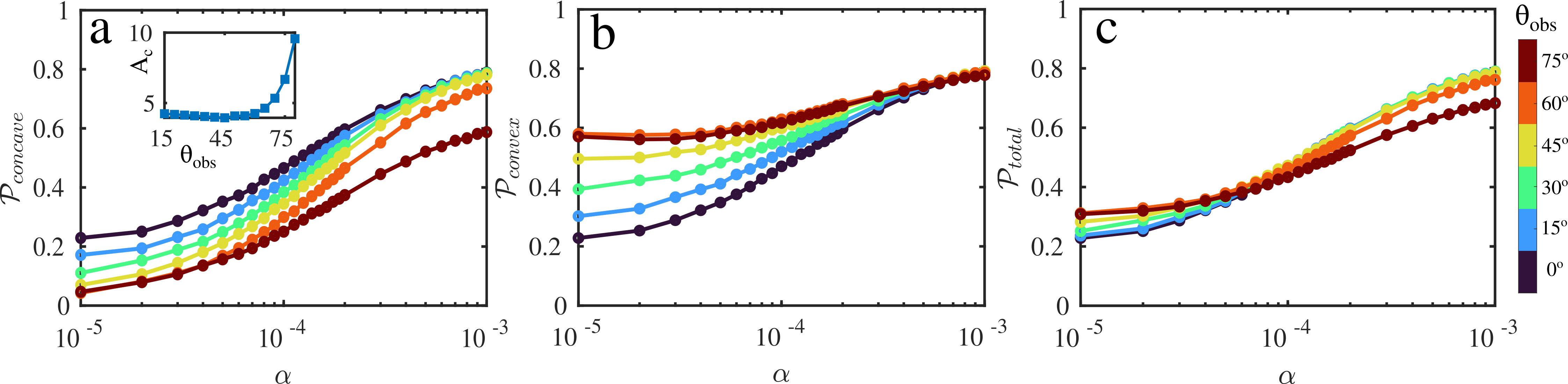}
 \caption{\textit{ Dependence of the available cavities for the passive particles to flow on the activity and geometry}.\textbf{(a)} The fraction of open concave cavities, $\mathcal{P}_{concave}$. The inset contains the clogging area, $A_c$, as a function of $\theta_{obs}$. Details on how to estimate the clogging area can be found in \textbf{Fig.S2} \textbf{(b)} The fraction of open convex cavities $\mathcal{P}_{convex}$, and \textbf{(c)} the total fraction of open cavities, $\mathcal{P}_{total}$ as functions of the tumbling rate, $\alpha$ and the angle of the obstacles, $\theta_{obs}$}
 \label{fgr:Clogging}
\end{figure*}

After showing that the first-order model captures the sorting dynamics upon variations in the geometry and the activity, aiming to understand their effects on it, we study the dependence of the kinetic parameters of the model on $\alpha$ and $\theta_{obs}$.
The results in Figure~\ref{fgr:KineticConstants} show, $\mathcal{F}_{bottom}$ (panel a) and $\tau_d$ (panel b) as a  function of $\alpha$ and $\theta_{obs}$. The rate constants can be used to calculate the fraction of passive particles $\mathcal{F}_{bottom}$ at steady state, as well as $\tau_d$. 
The fraction $\langle \mathcal{F}_{bottom} \rangle$ displays a maximum for all $\theta_{obs}$, in the range $1\times10^{-4}\leq\alpha\leq2\times10^{-4}$, with only a minor dependence on $\theta_{obs}$. The global maximum, obtained for $\theta_{obs} = 75\mycirc$, is approximately $\mathcal{F}_{bottom} =0.92$, meaning that the active-passive mixture initially placed at the top compartment is almost completely separated. This represents an improvement of around $30\%$ compared to the passive-passive case, where only steric interactions play a role ($\mathcal{F}_{bottom} \approx 0.70$; \textbf{See Fig. S4b}). When plotting $\langle \mathcal{F}_{bottom} \rangle$ as a function of the persistence length, $l_p = \frac{F_a dt}{\alpha\gamma_1}$,
as presented in the inset of Fig.~\ref{fgr:KineticConstants}a, we observe that the maximum takes place for the range $5\sigma_{22}\leq l_p \leq 10\sigma_{22}$, which is compatible with the length of the obstacles $\mathcal{L} = 3\sigma_{11} = 6\sigma_{22}$. This suggests that the maximum sorting is achieved when active particles move a distance of order $\mathcal{L}$ before tumbling.

As mentioned above, the position of the maximum slightly depends  on $\theta_{obs}$. However, the curves overlap for all values of the angle. Thus, we may conclude that the steady state partitioning of passive particles between the microchannel's compartments is largely independent of the shape of the intervening partition. This shape, however, affects the speed at which the steady state is reached. 
Figure~\ref{fgr:KineticConstants}b shows that the time constant $\tau_d$ decreases monotonically with increasing $\theta_{obs}$ for all $\alpha$, and displays a flat minimum in the range $1\times10^{-4}\leq \alpha \leq 4\times10^{-4}$ for all $\theta_{obs}$. This corresponds to the range of $\alpha$'s in which maximum sorting is achieved. Thus, selecting a tumbling rate in the range $1\times10^{-4}\leq \alpha \leq 4\times10^{-4}$ and $\theta_{obs} = 75\mycirc$ provides both maximum sorting and the fastest dynamics. 


The angle $\theta_{obs}$ influences the timescale $\tau_d$ both directly and indirectly. The direct effect has a geometric origin: the fraction of boundary that the passive particles can use to transit between top and bottom halves, depends on $\theta_{obs}$ as $\nu(\theta_{obs}) = \Delta/\left(\mathcal{L}\cos{\theta_{obs}} + \sigma_{33} + \Delta\right)$.
%
%
%
The timescale $\tau_d(\theta_{obs},\alpha)$ can therefore be written as 
\begin{align}
\tau_d(\theta_{obs},\alpha) &= \tau_d(0,\alpha)\frac{\nu(0)}{\nu(\theta_{obs})}(1+f(\theta_{obs},\alpha))\\
&=\tau_d(0,\alpha)\left[\frac{\mathcal{L}+ \sigma_{33} + \Delta}{\mathcal{L}\cos{\theta_{obs}} + \sigma_{33} + \Delta}\right](1+f(\theta_{obs},\alpha)),
\label{eq:taud}
\end{align}
where the function $f(\theta_{obs},\alpha)$ includes all the non-geometrical contributions to the difference between $\tau_d(\theta_{obs},\alpha)$ and $\tau_d(0,\alpha)$.
%
%
The inset in Fig.~\ref{fgr:KineticConstants}b shows the values of $f(\theta_{obs},\alpha)$ from simulations. We see that, for the set of parameters explored here, {$f(\theta_{obs},\alpha)\simeq0.0542\pm 0.1157$}.
This is a measure of the indirect influence of $\theta_{obs}$ and $\alpha$ on $\tau_d$ beyond the case $\theta_{obs}=0$, through their effect on the dynamics of active particles.

One of the main ways by which active particles influence transport of  passive ones is by blocking the gaps, 
hindering the flow of passive particles. To quantify this effect, we calculate the fraction of open gaps as a function of the tumbling rate of active particles. The results are presented in Fig.~\ref{fgr:Clogging}. Note that the steady-state distribution of the active particles within the top compartment of the microchannel is reached in a much shorter time scale than that of the system, which is determined by the steady distribution of passive particles within the microchannel (\textbf{see Fig.S5}), and thus the steady-state fraction of open gaps we are presenting also holds  for the transient of the system when a downward net flow of passive particles takes place.
For $\theta_{obs} = 0$, there is no distinction between concave and convex gaps, and thus the curves corresponding to this angle in Figs.~\ref{fgr:Clogging}a,b,c are the same. These curves set respectively the upper/lower limits of the fractions $\mathcal{P}_{concave}$ and $\mathcal{P}_{convex}$ for the other values of $\theta_{obs}$.
%
As we see in Fig.~\ref{fgr:Clogging}, both $\mathcal{P}_{concave}$ and $\mathcal{P}_{convex}$ are monotonically increasing functions of the tumbling rate $\alpha$, showing that active particles with shorter persistence always obstruct the gaps less than long-persistence ones, regardless of the value of $\theta_{obs}$. Interestingly, all curves show inflection points for $1\times10^{-4}\leq\alpha\leq 2\times10^{-4}$, marking a transition from a regime in which the gaps are blocked most of the times, to another where the gaps are mostly free. This range corresponds  with maximum sorting (see Fig \ref{fgr:KineticConstants}a), highlighting the connection between gaps' dynamics and passive particle segregation.
%
From this behavior, we may already infer that the most effective sorting relies more on the way  active particles enhance the transport of passive particles than on the availability of gaps for the passive particles to flow towards the bottom compartment. 

\begin{figure}[t]
 \centering
 \includegraphics[width=7.0cm]{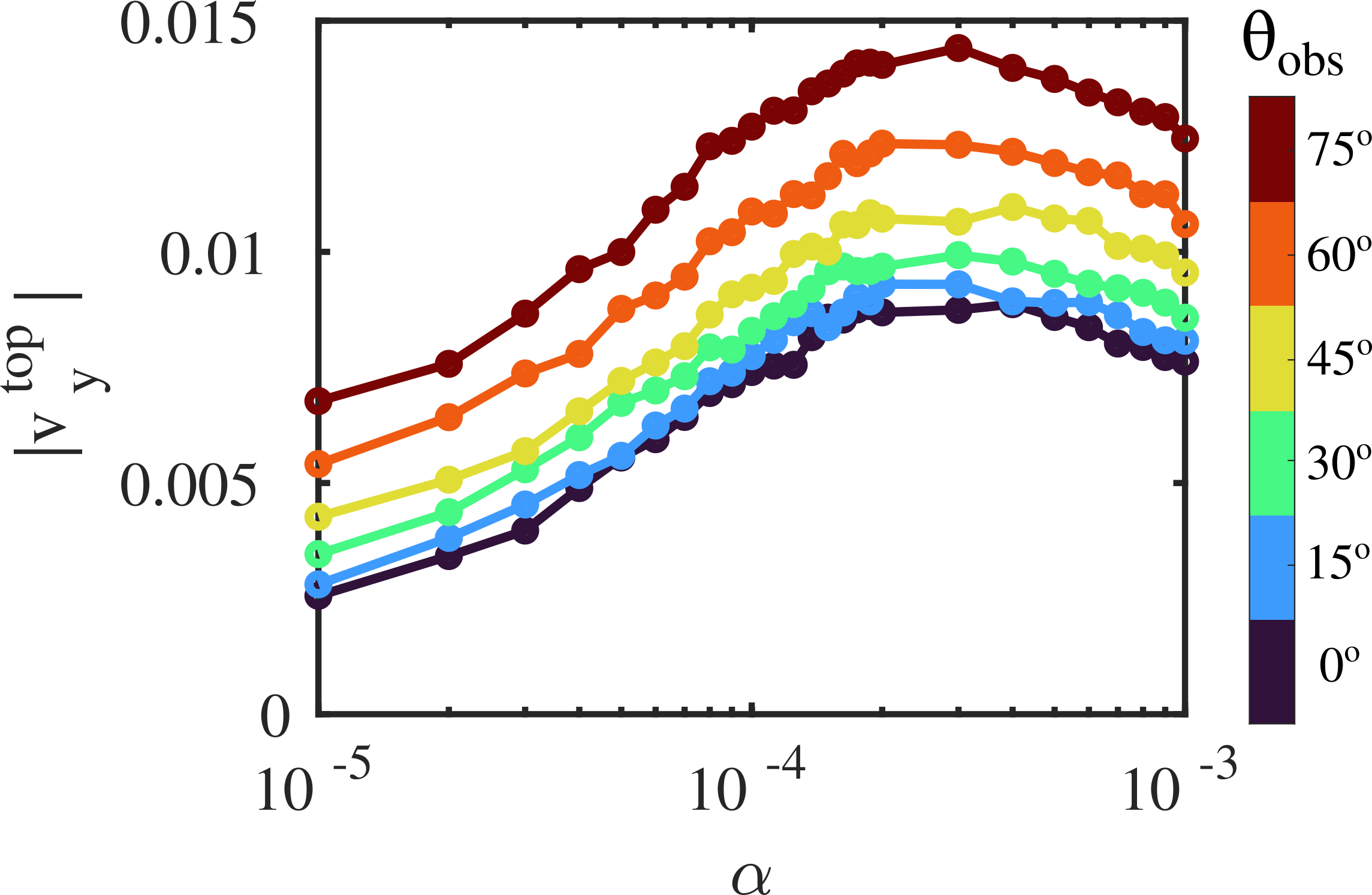}
 \caption{\textit{Drift velocity in the $y$-direction induced by the concomitant effect of activity and confinement} The drift velocity in the top compartment of the microchannel as a function of the tumbling rate, $\alpha$, and the angle of the obstacles, $\theta_{obs}$.}
 \label{fgr:DriftY}
\end{figure}

Figure~\ref{fgr:Clogging}a shows that, in contrast to the dependence on tumbling rate, $\mathcal{P}_{concave}$ is a monotonically decreasing function of $\theta_{obs}$, a behaviour correlated with the corresponding increase in the clogging area $A_{c}$ (Fig.~\ref{fgr:Clogging}a, inset). This is to be expected, as increasing $\theta_{obs}$ -and therefore $A_{c}$- leads to a larger set of configurations for single active particles to be able to clog concave pores.
This is compounded by the tendency of active particles to accumulate at the concave regions, forming a multilayered structure that becomes more pronounced as $\theta_{obs}$ increases and $\alpha$ decreases \textbf{(See Fig.~S8}).
%
%
%
The situation is opposite for $\mathcal{P}_{convex}$, which is a monotonically increasing function of both $\alpha$ and $\theta_{obs}$ (Fig.~\ref{fgr:Clogging}b), despite the fact that the clogging area of convex gaps is independent of $\theta_{obs}$.
The cause is the increase in the fraction of active particles within the concave regions as $\theta_{obs}$ increases. On the one hand,  this leads  to fewer active particles available to close the convex gaps. On the other, the higher concentration of active particles in concave regions leads to more frequent interactions capable of removing active particles from within the clogging areas of convex regions.
Altogether this leads to a relative increase in $\mathcal{P}_{convex}$ with  $\theta_{obs}$.  
It is interesting to notice that, for $\alpha\gtrsim3\times10^{-3}$, $\mathcal{P}_{convex}$ ceases to depend on either $\alpha$ or $\theta_{obs}$. 
This is due to the rapid change in orientation and the fact that, contrary to the concave case, active particles do not form layered structures around convex gaps. Finally, Figure~\ref{fgr:Clogging}c shows that, contrary to $\mathcal{P}_{concave}$ and $\mathcal{P}_{convex}$, the total fraction of open gaps, $\mathcal{P}_{total} = (\mathcal{P}_{concave} + \mathcal{P}_{convex})/2$ is largely independent of $\theta_{obs}$ (Fig.~\ref{fgr:Clogging}c). Deviations become more prominent for  $\theta_{obs} = 75\mycirc$, mostly due to the increased tendency of active particles to form  multilayered structures at concave gaps. The small deviations we do observe can be either positive or negative depending on $\alpha$, with a crossover around $\alpha\simeq6\times 10^{-5}$, corresponding to a persistence length, $l_p \approx 17\sigma_{22}$, which correlates with the size of the top compartment of the microchannel, $L_y/2 = 20\sigma_{22}$. An alternative way of understanding $\mathcal{P}_{convex}$ and $\mathcal{P}_{cocave}$ is by considering the average time convex and concave gaps remain clogged ($t^c_{concave}$ and $t^c_{convex}$) and unclogged ($t^u_{concave}$ and $t^u_{convex}$). The expressions are $\mathcal{P}_{concave} = \frac{t^u_{concave}}{t^u_{concave} + t^c_{concave}}$ and $\mathcal{P}_{convex} = \frac{t^u_{convex}}{t^u_{convex} + t^c_{convex}}$, as shown in \textbf{Fig. S11}.

\begin{figure}[bt]
 \centering
 \includegraphics[width=7.0cm]{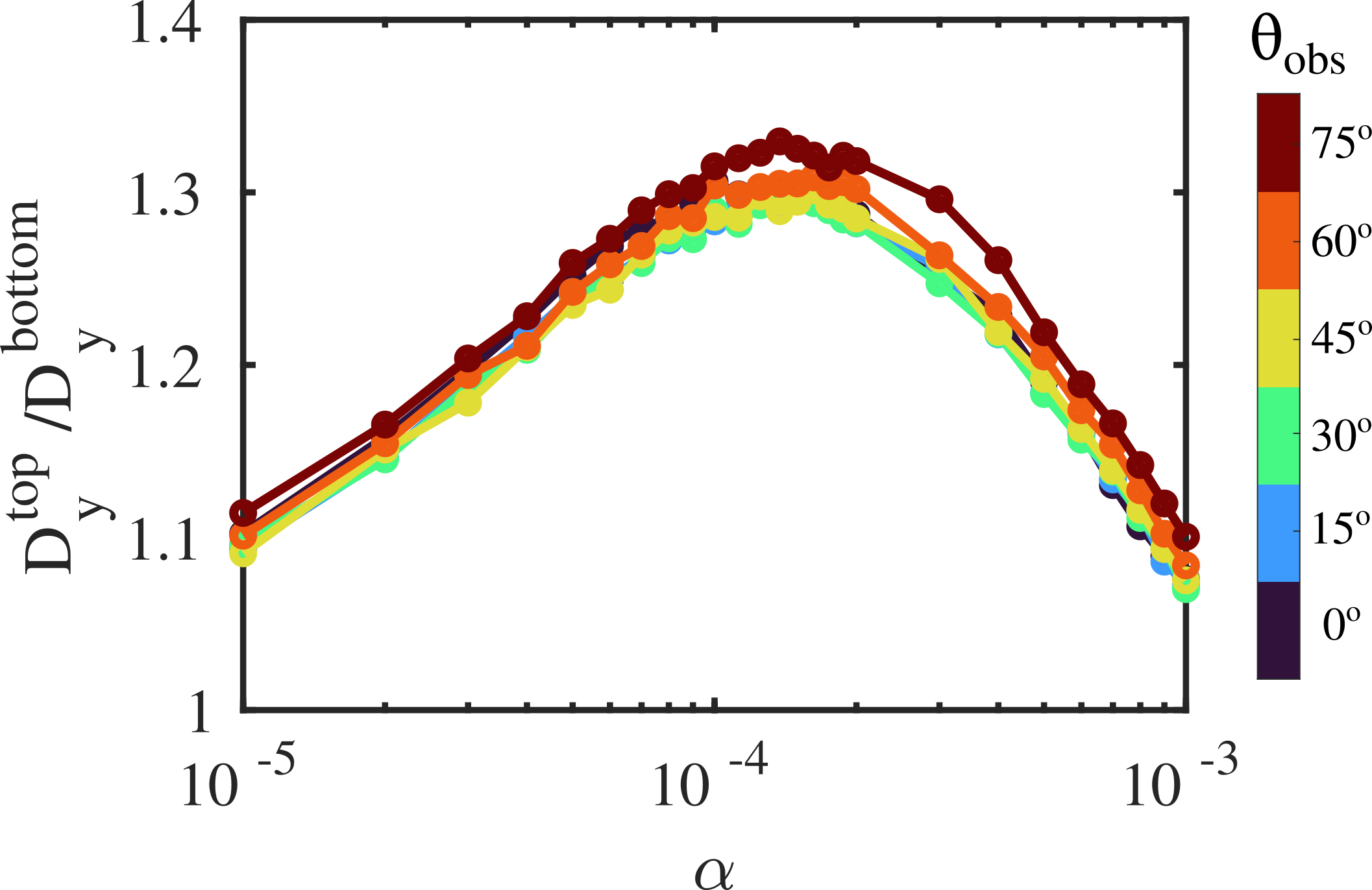}
 \caption{\textit{Enhancement of the diffusion coefficient of passive particles in the active bath}.The ratio between the diffusion coefficients in the top and bottom compartments along the $y$ direction, $D_y^{top}/D_y^{bottom}$, as a function of the tumbling rate, $\alpha$, and the angle of the obstacles, $\theta_{obs}$.}
 \label{fgr:DiffusionY}
\end{figure}

The enhancement of the transport properties of passive particles within the active bath is a direct consequence of the encounters between active and passive particles. 
It can be characterised by looking at the time evolution of the probability distribution function (PDF) of passive particles displacements. We will focus here on transport along the $y$-direction, which is the one relevant for particle sorting. More information on the dynamics along the $x$-direction -the channel axis- can be found in the \textbf{SI}. We calculate the PDF of displacements $\Delta y$ within a time interval of $\Delta t$, in the top and the bottom sections, $\text{P}_{top/bottom}(\Delta y, \Delta t)$ (\textbf{Fig.~S7}) and extract both the first and second moments of these distribution as a function of time and position. 
%
The passive particles' effective diffusivity in the top/bottom halves, $D_y^{top/bottom}$, can be calculated from the variance of $\text{P}_{top/bottom}(\Delta y, \Delta t)$, which is the (de-drifted) mean-squared displacement of the particles (\textbf{Fig.~S7}). This should be considered at times intermediate between the typical interval between interactions with active particles  and the time required to diffuse a distance of order $L_y$. Figure~\ref{fgr:DiffusionY} shows the ratio $D_y^{top}/D_y^{bottom}$, where the bottom dynamics is due to thermal diffusion of confined passive particles. 
We observe that the ratio of diffusion coefficients in the $y$-direction reaches a maximum in the range $1\times10^{-4}\leq \alpha \leq 2\times10^{-4}$, which is compatible with those reported for maximum sorting (Fig.~\ref{fgr:KineticConstants}a), fastest time response (Fig.~\ref{fgr:KineticConstants}b) and inflection point of $\mathcal{P}_{total}$ (Fig.~\ref{fgr:Clogging}c). This confirms that the best separation performance of the system correlates with the maximization of the activity-induced diffusivity along the $y$-direction within the top compartment of the microchannel.


Besides the variations in diffusivity, the interactions with active particles in the top section generate a surprising net drift of passive particles, $v_y^{top}$, directed downward. This drift can be derived from the first moment of $\text{P}_{top}(\Delta y, \Delta t)$ and it is a consequence of the asymmetry between the two boundaries confining the active particles, as confirmed by simulations (\textbf{Fig.~S6,S7}). Figure~\ref{fgr:DriftY} shows the dependence of $v_y^{top}$ on both $\alpha$ and $\theta_{obs}$. It reaches a maximum for $\theta_{obs}=75\mycirc$ and $\alpha \in [1\times10^{-4}, 2\times10^-{-4}]$, corresponding to the maximum sorting and  shortest time constant $\tau_d$. The active particles not only enhance the diffusion coefficient of the passive particles but also induce an active drift that -as we will see- plays a crucial role in enhancing the flow of passive particles through the gaps. 

\subsection{A minimal one-dimensional Advection-Diffusion model}


The effect of activity on the segregation dynamics of passive particles can be rationalised following a simple Advection-Diffusion (AD) model similar to the one proposed in \cite{williams2022confinement}. We consider a one-dimensional system in the domain $y \in [0, \rho L_y]$, with the factor $\rho$ fixed {\it a posteriori} as indicated below. Within this domain, the concentration of passive particles, $c(y,t)$, evolves in time according to
\begin{equation}
\frac{\partial c}{\partial t} = \frac{\partial}{\partial y} \left[\frac{\partial}{\partial y}\left ( D(y)c\right) \right] - \frac{\partial}{\partial y} \left[ v(y) c \right].
\label{e:AdvectionDiffusion}
\end{equation}
Here $D(y)$ is the position-dependent diffusion coefficient and $v(y)$ is the effective drift velocity of the passive particles. As mentioned earlier, their values in the top/bottom sections, $D_\text{top/bottom}$ and $v_\text{top/bottom}$, are calculated from the first two moments of the displacements PDFs for the passive particles. In the following we will always consider $v_\text{bottom}=0$. The two values will then be heuristically interpolated using the sigmoids.
\begin{equation}
D(y) = D_\text{bottom} + \frac{D_\text{top} - D_\text{bottom}}{1 + \exp(-k(y - L_y/2))},
\label{e:Dsigmoid}
\end{equation}
\begin{equation}
v(y) = \frac{v_{top}}{1 + \exp(-k(y - L_y/2))}.
\label{e:vsigmoid}    
\end{equation}
with $k=100$.
Particle concentration is ensured through no-flux boundary conditions at both ends of the domain:
\begin{align}
\left. - \frac{\partial}{\partial y}\left(D(y)c\right)+ v(y)c \right|_{y=0,\rho L_y} &= 0. 
\label{e:BCs}
\end{align}
\begin{figure*}[h!]
 \centering
 \includegraphics[width=18.0cm]{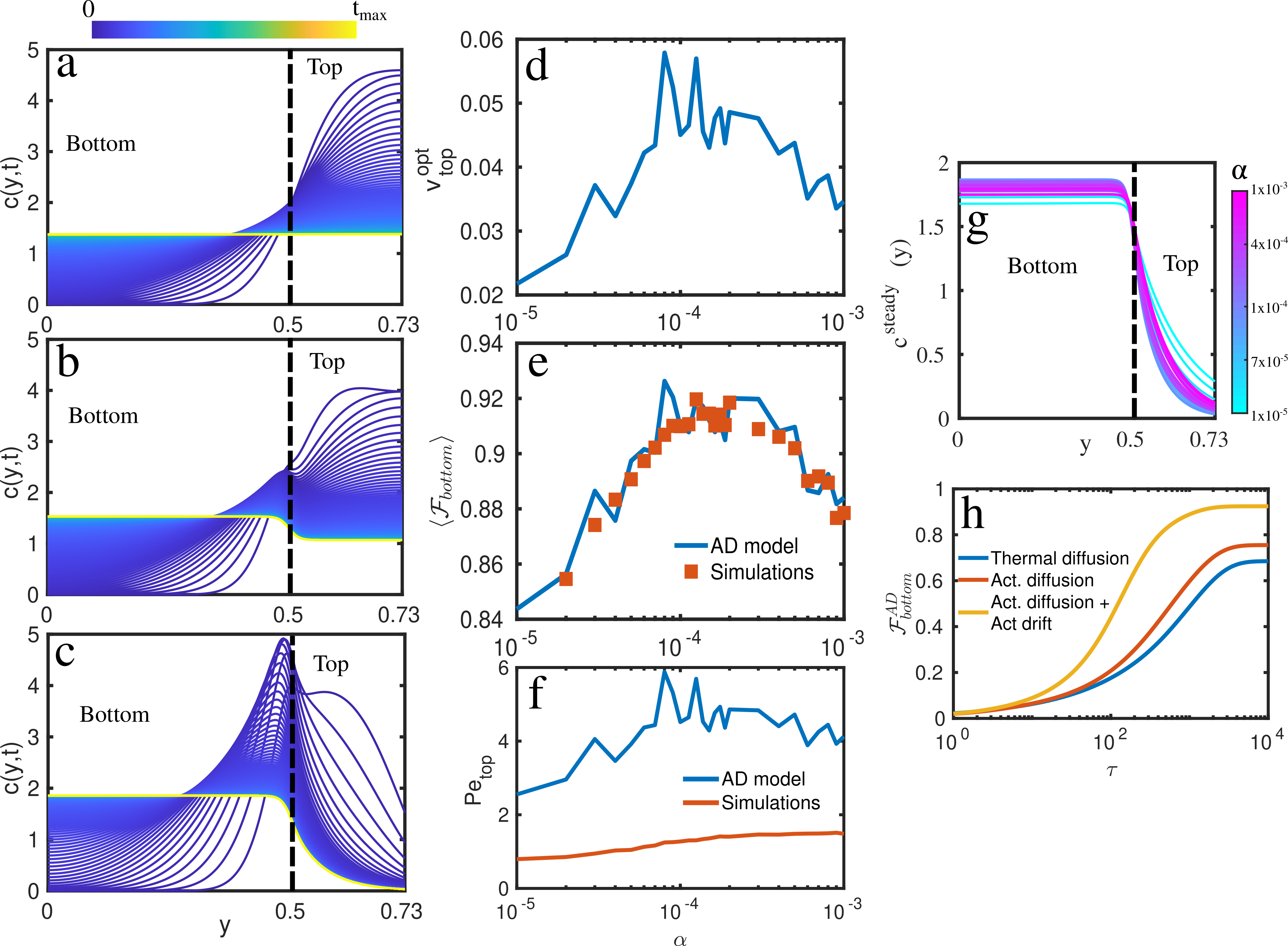}
 \caption{\textit{Advection-Diffusion model}. In all the results $\rho = 0.73$ and $\theta_{obs} = 75^{\mycirc}$. The time evolution of the concentration profile: \textbf{(a)} pure thermal diffusion case ($D_{top}/D_{bottom} = 0.75$,  $v_{top} = 0$, $\partial D/\partial y = 0$); \textbf{(b)} thermal diffusion + active diffusion ($D_{top}/D_{bottom} = 1.30$, $v_{0} = 0$, $\partial D/\partial y \neq 0$); \textbf{(c)} thermal diffusion + active diffusion + active advection for $v_{top} = -0.056$, $\partial D/ \partial y \neq 0$. \textbf{(d)} The best fitted values of the top drift velocity, $v_{top}^{opt}$, as a function of the tumbling rate, $\alpha$. \textbf{(e)} Comparison between the steady state concentration of passive particles in the bottom compartment calculated from simulations and the A-D model. \textbf{(f)} Comparison of the P\'eclet number in the top compartment calculated in the simulations and the A-D model according to $Pe_{top} = \frac{L_y|v_{top}|}{2D_{top}}$. \textbf{(g)} The steady state concentration profiles obtained in the A-D model with the best fitted drifts for all the values of $\alpha$ considered. \textbf{(h)} The time response of the concentration of passive particles in the bottom compartment, calculated from the A-D model,  $\mathcal{F}^{AD}_{bottom}$, following first order dynamics as in the simulations.}
 \label{fgr:A-D_model}
\end{figure*}
The initial concentration field $c(y,0)$ is localised in the top part of the container and it is normalised to 1.
The time evolution of the concentration profile (Fig.~\ref{fgr:A-D_model}a,b,c) is calculated with a conservative explicit finite difference method \cite{leveque2002finite,morton2005numerical}.

When both type 1 and type 2 particles are passive, corresponding to  $D_{top}/D_{bottom} = 0.75$ , 
simulations show that $\langle \mathcal{F}_{bottom}\rangle= 0.69$ (See \textbf{Fig. S4 b and g}) A smaller diffusion coefficient in the top in the passive-passive case is related to the space occupied by the big passive particles that reduce the available space for the small passive particles to diffuse. In this case, although $D_{top}\neq D_{bottom}$, we assume that $\partial D/ \partial y = 0$ since there is no activity in the top compartment (see \textbf{SI}, equations \textbf{S1-S6}). In the active-passive case, for each value of the tumbling rate $\alpha$ we start by fixing $D_{bottom}=1$ while $D_{top}$ is set to the value obtained from simulations (see Fig.~\ref{fgr:DiffusionY}). In this case, $\partial D / \partial y$ is given by equation \textbf{S4} of the \textbf{SI}. Note that we also rescale $L_y = 1.0$, which jointly with the $D_{bottom} = 1.0$ define the time scale $\tau_{AD} = L_y^2/D^{bottom}_y$ We then determine the advection velocity $v_{top}$ that minimizes the discrepancy between the steady state concentration in the bottom side of the compartment evaluated with the model 
and the reference value $\langle \mathcal{F}_{bottom}(\alpha)\rangle$ obtained from simulations. 
Figure~\ref{fgr:A-D_model}d shows that, for an appropriate choice of $v_{top}$, the model in eq.~\ref{e:AdvectionDiffusion} reproduces well the steady state separation observed in simulations. The key nondimensional  parameter here is the ratio $\text{Pe}_{top}=|L_yv_{top}/2D_{top}|$, which is akin to a P\'eclet number for the active displacements of the type 2 particles in the top container. Comparing the values of $\text{Pe}_{top}$ from simulations and for the advection-diffusion model (Fig.~\ref{fgr:A-D_model}e) we see a discrepancy by a factor $\lesssim 4.5$, which is appropriate for a minimal model that distills
the key dynamics of a complex process in a simple and intuitive way.
%
Having determined the model parameters that best represent the steady state of the simulation, we can go back to the full A-D model and use it to explore which features of the active displacements of passive particles contribute most to particle segregation. We will consider here the case corresponding to  maximum segregation ($\alpha=1.25\times10^{-4}$, $\theta_{obs}=75^{\circ}$), leading to $\langle F_{bottom}\rangle=0.92$ at steady state. Figures~\ref{fgr:A-D_model}a, b, c show the time evolution of $c(y,t)$ from eq.~\ref{e:AdvectionDiffusion} for the cases of thermal diffusion ($v(y) = 0$, $D_{top}/D_{bottom} = 0.75$, $\partial D / \partial y = 0$), active diffusion ($v(y) = 0$, $D_{top}/D_{bottom} = 1.30$), active diffusion and active advection ($v_{top} = -0.056$, $D_{top}/D_{bottom} = 1.30$).
Although in all three cases the time evolution of $\langle \mathcal{F}_{bottom}\rangle$ follows a first order kinetic (See Fig \ref{fgr:A-D_model} h), the levels of passive particles segregation achieved at steady state are substantially different.
In the first case, at steady state we obtain $\mathcal{F}_{bottom}=0.682$, which agrees with the simulation results of the passive-passive case as a consequence of our choice of $\rho$. 
%
%
Upon considering the space-dependent diffusion profile, $\mathcal{F}_{bottom}$ is increased to 0.732 due to the flux induced by the diffusivity gradient. 
Finally, considering the full model with $v_{top} = -0.056$, leads to $\mathcal{F}_{bottom}=0.921$. This last value is in good agreement with the sorting measured in simulations.
These results suggest that the main physical origin of the improved sorting compared with the passive-passive case is the advection induced by the active bath rather than the difference in diffusivity between top and bottom containers. 
\textcolor{black}{This is confirmed also by simulations in which the obstacles are replaced by a simple flat reflective wall affecting only the active particles (\textbf{Fig.~S6}). Although the system generates a non-uniform profile of diffusivity, the symmetry between the walls seen by the active particles implies no net advection ($v(y)=0$).
The results show a trapping efficiency that varies only within $\sim 5\%$ of the passive-passive case, in line with the findings above.}
%

The role of advection can be rationalised further by taking a closer look at the steady state solution $c(y)$ of the AD model. 
\begin{equation}
c(y)=\frac{A}{D(y)}e^{\int^y\frac{v(y)}{D(y)}dy},
\end{equation}
where $A$ is a normalisation constant. The fraction $\mathcal{F}_{bottom}$ will therefore be equal to 
\begin{equation}
\mathcal{F}_{bottom}^{AD} = \frac{\int_{0}^{L_y/2}c(y)\,dy}{\int_{0}^{\rho L_y}c(y)\,dy}\simeq \frac{1}{1+\frac{2D_{bottom}}{v\,L_y}\left(1-e^{-\rho Pe_{top}}\right)}.
\label{eq:ADFbottom}
\end{equation}
Here the rightmost expression has been calculated by approximating the profiles in eqs.~\ref{e:Dsigmoid},\ref{e:vsigmoid} as piecewise constant. Using the simulation results in Figs.~\ref{fgr:DiffusionY},\ref{fgr:DriftY} it is easy to show that, in the present system, the curve $\mathcal{F}_{bottom}^{AD}(\alpha)$ is almost entirely dictated by $v(\alpha)$ (\textbf{Fig.~S10, Supplementary Material}). 
%
The peak in separation performance occurs therefore as a consequence of a peak in the advection velocity. In turn, this peak corresponds to the range of tumbling rates $\alpha$ for which the persistence length of the active particles is of the order of the length of the obstacles, $\mathcal{L} = 6\sigma_{22}$, as well as to the inflection point in the fraction of open gaps at  steady state (Fig.~\ref{fgr:Clogging}c), which separates the regime where the gaps are mostly closed (lower $\alpha$) from the one where they are mostly open (larger $\alpha$).
%
Altogether, the peak in advection speed can be related to a balance between two competing effects. On the one hand, increasing the tumbling rate increases the rate at which active and passive particles interact, due to a lower accumulation of active particles at the boundaries. On the other hand, decreasing the tumbling rate increases the expected displacement of passive particles due to their interaction with active ones. The peak happens at an optimal value that balances frequency of interactions and active transport per interaction event. The advection speed in the top compartment induced by the concomitant effect of confinement and activity constitutes the principle of an \textit{active-pumping} mechanism that facilitates the separation of the passive species and might be exploited in technological applications like cargo delivery and micro separators. 


\section{Conclusions}
\label{Conclusions}

We have studied by means of computer simulations the sorting dynamics of a two-dimensional active-passive mixture within designed microchannels comprising confinement in the $y$ direction and funnel-like obstacles in their central region. The funnels feature gaps that only allow the passage of passive particles, so they behave like a semi-permeable membrane to that effect. The simulations were carried out in the overdamped regime to neglect inertial effects but solving the complete Langevin's equation for the translational dynamics for the sake of generality. The active particles followed a run-and-tumble motion for the rotational dynamics. We have observed that the dynamics of the sorting process follows a first-order kinetics model, allowing us to study the dependence of the kinetic constants with the activity and the geometry of the microchannels.

We have observed that there is a small range of tumbling rates that guarantee the highest fraction of passive particles separated from the mixture and the fastest dynamics. The existence of the  most favorable range can be explained in terms of an interplay of maintaining a reasonable fraction of open gaps for the passive particles to flow (See Fig. \ref{fgr:Clogging} c), maximizing the active kicks experienced by the passive particles, that in turn enhance the diffusion coefficient of the passive particles (See Fig. \ref{fgr:DiffusionY}) and induce a negative advective drift (See Fig. \ref{fgr:DriftY} and \textbf{ Fig. S7}) and the geometry of the funnels that allows the maximum relative space for the flow of passive particles (See Fig. \ref{fgr:KineticConstants}b). Solving a minimal Advection-Diffusion model, we determined that the flow of passive particles towards the bottom compartment and their subsequent sorting requires a negative active drift velocity in the top compartment. Such sorting is neither observed only by virtue of the difference of the diffusivity between compartments nor the active diffusion associated to the drift term as the gradient of the diffusivity (See Fig. \ref{fgr:A-D_model}).

Finally, we rationalise the position and the existence of the peak in the advective speed, and in turn the system's peak performance, by the balancing of two competing effects determined by the the tumbling rate: the increase of the frequency of encounters between active and passive particles (number of active-kicks)as $\alpha$ increases and the active transport per interaction event (the length of active displacements) as alpha decreases. The interplay of these effects are the basis of an \textit{active-pumping} mechanism that might be useful in the development of cargo delivery systems and separators of micro-contaminants 

Future work will consider including  hydrodynamic interactions to understand its role on the sorting dynamics and an experimental realization of the system using mixtures of \textit{C. reinhardtii} and passive beads confined in microfluidics devices with funnel-like geometries. The results presented here might help developing strategies based on microorganisms' taxis to induce a synchronized motion that might lead to a further increase of the advective velocity ( synchronous \textit{active-pumping}) improving even further the sorting of passive particles in the designed microchannels. Additionally it might be interesting to study the effects of $\chi_{active}$ on the system's dynamics to discern the role of the multi-layered structure formed on the gaps in their clogging.

\section*{Conflicts of interest}
There are no conflicts to declare.

\section*{Data Availability Statement}
The data that support the findings of this study are available from the corresponding author upon reasonable request.

\section*{Acknowledgements}

M.P. acknowledges funding from MINECO through projects CNS2022-135975 and PID2023-146578NB-I00, and acknowledges that the fact that IMEDEA is an accredited ‘María de Maeztu Excellence Unit’ (grant CEX2021-001198, funded by MCIN/AEI/10.13039/ 501100011033). C.V. acknowledges fundings  IHRC22/00002 and PID2022-140407NB-C21 from MINECO. H.S and C.V acknowledge the received funding from the European Union’s Horizon research and innovation programme under the Marie Skłodowska-Curie grant agreement No 101108868 (BIOMICAR). H.S. acknowledges enriching discussions with José Martín-Roca, Juan Pablo Miranda López and Sujeet Kumar Choudhary. I.P. acknowledges support from Ministerio de Ciencia, Innovaci\'on y Universidades MCIU/AEI/FEDER for financial support under grant agreement PID2021-126570NB-100 AEI/FEDER-EU, and Generalitat de Catalunya for financial support under Program Icrea Acad\`emia and project 2021SGR-673

\balance

\bibliography{BIOMICAR_SERNA}
\bibliographystyle{ieeetr}

\end{document}


\maketitle
\section{Langevin Dynamics vs Brownian Dynamics}
\begin{figure*}[h!]
 \centering
 \includegraphics[width=17.0cm]{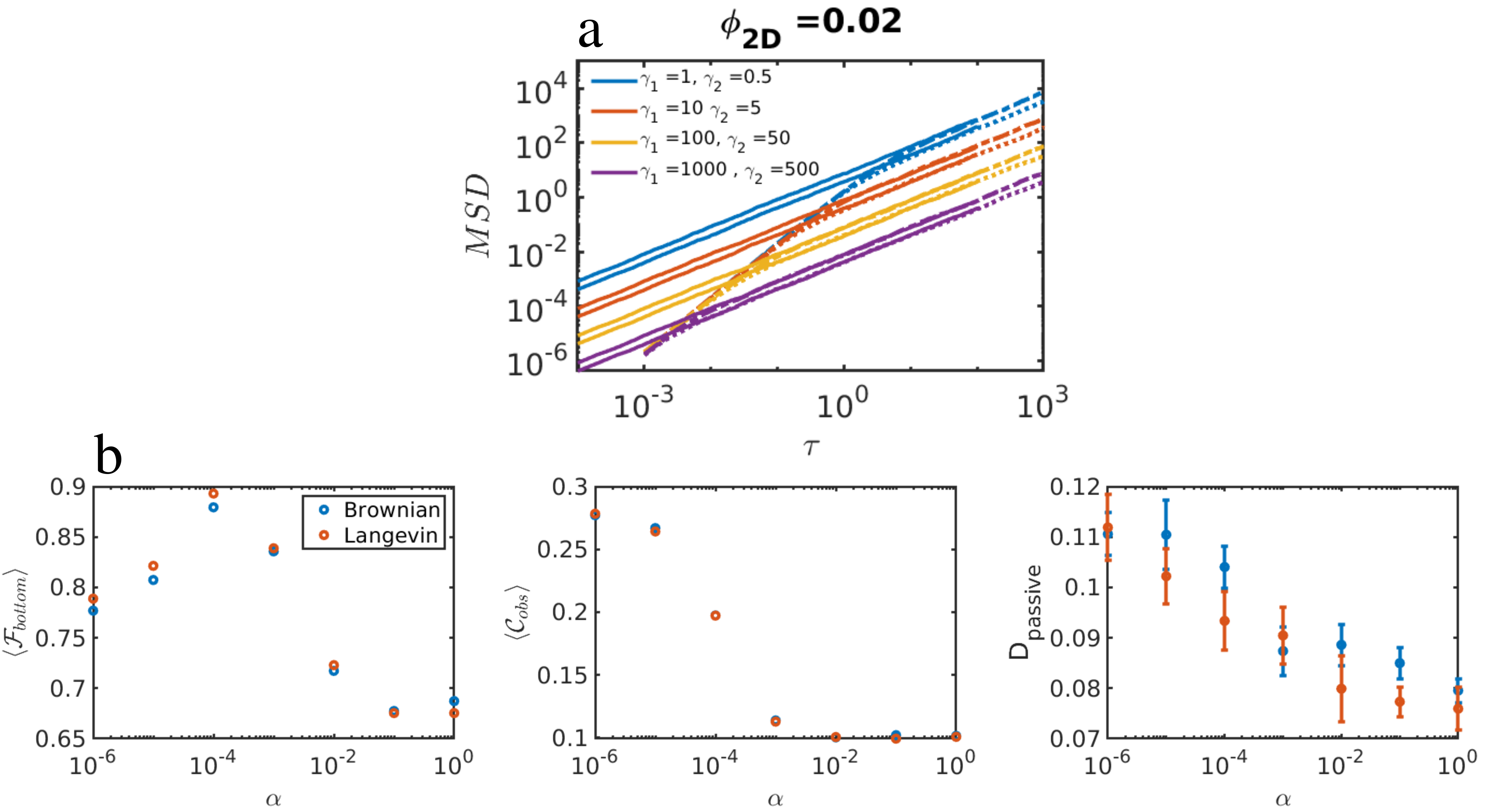}
 \caption{\textit{Comparisson between Langevin Dynamics (LD) and Brownian Dynamics (BD)}. \textbf{(a)} The mean squared displacement (MSD) of a passive-passive ($\sigma_{11} = 2$ and $\sigma_{22} = 1$) mixture in a very diluted case ($\phi_{2D} = 0.02$) when solving the complete Langevin's equation (LD) (dashed lines) and  the overdamped approximation of Brownian Dynamics (BD) (solid lines) for different sets of friction coefficients, $\gamma_1$ and $\gamma_2$. As described in the main text, the ratio $\gamma_1/\gamma_2 = \sigma_{11}/\sigma_{22} = 2$ as a consequence of Stokes' law. The MSD is given in $\sigma^2_{22}$ units. So, for each color, the higher $MSD$ corresponds to small particles and the lower to big particles. Both methods coincide for long times. In our simulations, we have chosen $\gamma_1 = 10$ and $\gamma_2$ = 5 (in red), which exhibits some ballistic regime only for $\tau < 1$, while the simulations we present in the main text run for $\tau = 20000$. With these values of the friction coefficients, we guarantee that particles move sufficiently to avoid extremely long simulations while making inertial effects less relevant. \textbf{(b)} Simulations of the active-passive system confined in the microchannels comparing LD and BD. The obstacles of the microchannel have a length $\mathcal{L} = 1.5\sigma_{11}$ and an angle $\theta_{obs} = 45^o$. The average fraction of passive particles in the bottom compartment of the microchannel at steady state, $\langle \mathcal{F}_{bottom}\rangle$ (left panel), the average fraction of accumulated active particles in the obstacles (funnels) region at steady state, $\langle \mathcal{C}_{obs}\rangle$ (central panel), and the diffusion coefficient of the passive particles, $D_{passive}$ (right panel), adopt very similar values with both methods when varying the tumbling rate, $\alpha$.}
 \label{fgr:LDvsBD}
\end{figure*}

\clearpage

\section{Estimation of the clogging area}

\begin{figure*}[h!]
 \centering
 \includegraphics[width=17.0cm]{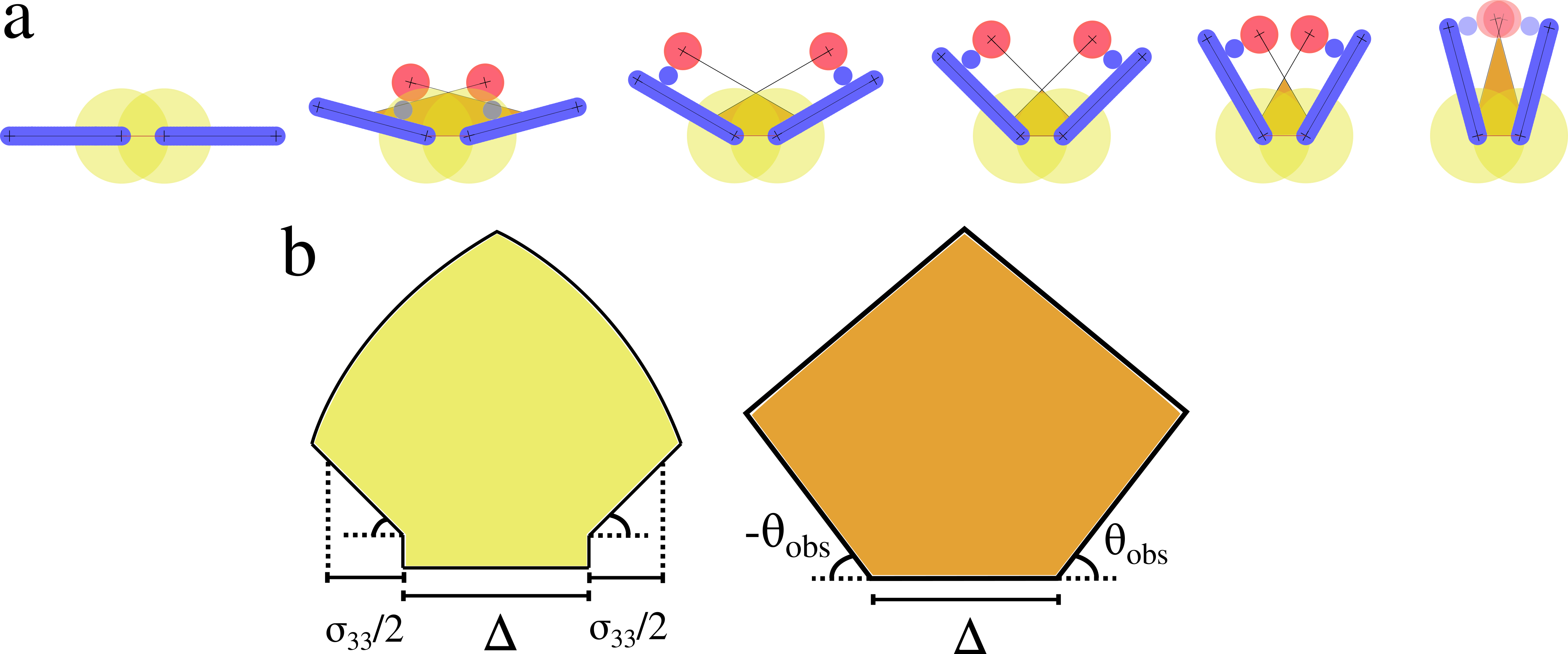}
 \caption{\textit{Clogging Areas of concave gaps for different $\theta_{obs}$}. The clogging criterion for concave gaps requires two sub-criteria: 1. For small angles $\theta_{obs}\leq30^o$, we need to use the intersection of the two yellow circles in \textbf{(a)} as the criterion. The radius of this circles is $R = \sigma_{33}/2 + \sigma_{22} + \sigma_{11}/2$. 2. For large angles ($\theta \geq 30^o$), the walls of the obstacles start to play a role since now an active particle close to the walls also avoid the passage of passive particles. In this case, we take into account the orange area below the two straight lines parallel to the obstacles as constructed in \textbf{(a)}, more details are given in the main text. Observe that for small angles, there are some positions inside the orange areas available for the active particles that still allow the passage of passive particles, this is why we need to use the 2 sub-criteria. In \textbf{(a)} we show how these two areas change as functions of $\theta_{obs}$. In \textbf{(b)}, schemes for the estimation of both areas are presented. The areas can be calculated with the information of the geometry and sizes of the particles involved in the simulations. Although they are not exact, the estimations serve well for our purpose as shown in \textbf{(a)}}
 \label{fgr:CloggingAreas}
\end{figure*}

\clearpage

\section{Independence of $\mathcal{P}_{total}$ with $\theta_{obs}$}

The relative independence of $\mathcal{P}_{total}$ on $\theta_{obs}$ can be understood as a consequence of the symmetry between concave and convex gaps. If we call $\mathcal{P}(\theta_{obs})$ the probability of a gap to be open when the obstacles are oriented at an angle $\theta_{obs}$, then the symmetric shape of the boundary in our simulations implies that $\mathcal{P}_{total}(\theta_{obs})=(\mathcal{P}(\theta_{obs})+\mathcal{P}(-\theta_{obs}))/2$. Here a positive/negative value of $\theta_{obs}$ indicates a concave/convex gap. As the function $\mathcal{P}_{total}(\theta_{obs})$ is even, its deviation from $\mathcal{P}_{total}(0)$ is at least $\mathcal{O}(\theta_{obs}^2)$. We therefore expect -and observe- only a weak dependence of $\mathcal{P}_{total}(\theta_{obs})$ on $\theta_{obs}$. 
The small deviations we do observe can be either positive or negative depending on $\alpha$, with a crossover around $\alpha\simeq6\times 10^{-5}$. 

The mild dependence of $\mathcal{P}_{total}$ on $\theta_{obs}$ does not imply that a convex gap is functionally equivalent to a concave one; rather, it indicates that when symmetric pairs of such gaps are considered, the enhancement in the flux of passive particles through one is mostly balanced by the reduction through the other.

It is interesting to note that, as a function of tumbling rate, all $\mathcal{P}_{total}(\theta_{obs}, \alpha)$ curves show an inflection point for $\alpha \in [1\times10^{-4},2\times10^{-4}]$, corresponding to $\mathcal{P}_{total} \simeq 0.50$. This clarifies the observed dependence of $k^-$ on $\alpha$ (\textbf{Fig.~S3b}), since one of the most important factors determining the flow of passive particles from the bottom compartment to the top is the fraction of open gaps. 

\begin{figure*}[h!]
 \centering
 \includegraphics[width=17.0cm]{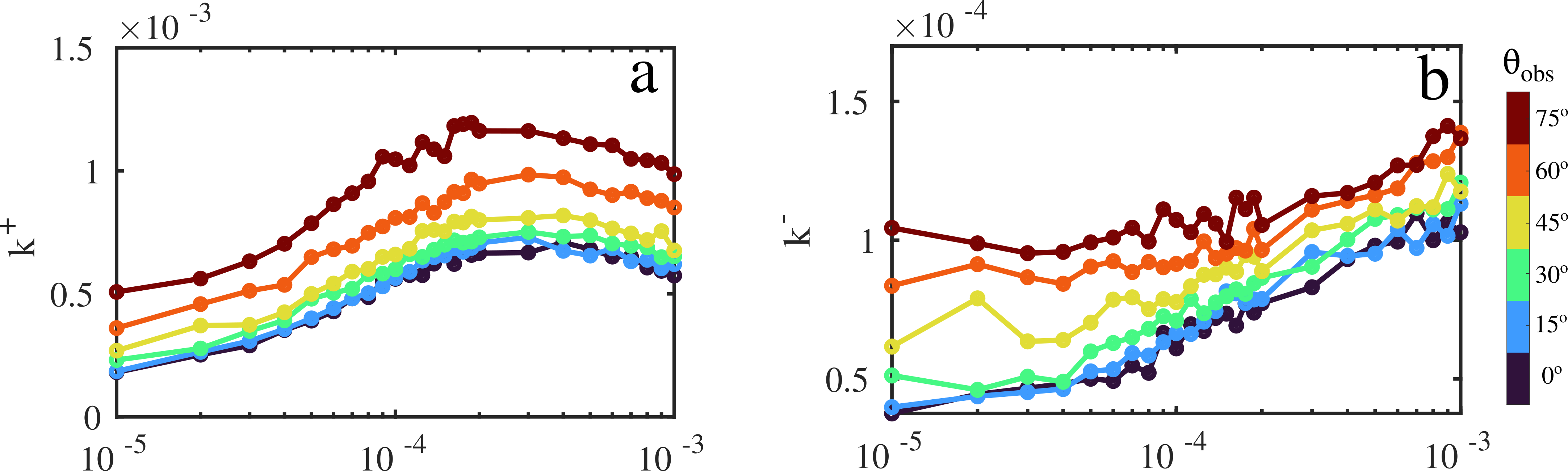}
 \caption{\textit{Kinetic constants}\textbf{(a)}$k^+$ and \textbf{(b)} $k^-$ as functions of the tumbling rate, $\alpha$, and the angle of obstacles, $\theta_{obs}$. $k^+$ determines the flow of passive particles from top to bottom and $k^-$ the flow of passive particles from bottom to top.} \label{fgr:Passive-Passive}
\end{figure*}

\clearpage

\section{Dynamics of the passive-passive case}

\begin{figure*}[h!]
 \centering
 \includegraphics[width=17.0cm]{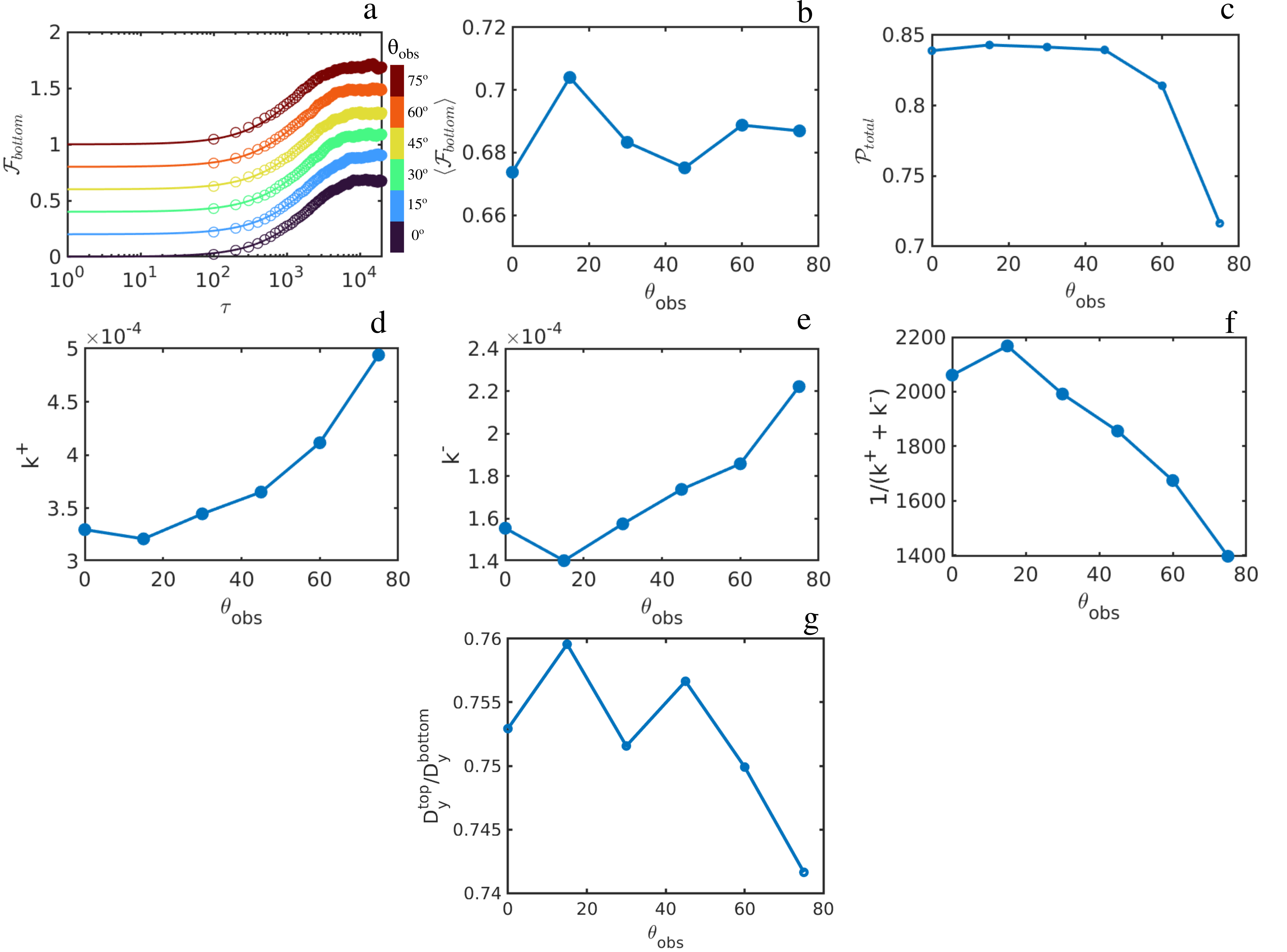}
 \caption{\textit{Dynamics of the passive-passive system in the microchannels}. All the parameters are the same as the system studied in the main text, with the only difference that type 1 and type 2 particles are now passive. \textbf{a} The markers are the simulation data, and the solid lines are the fits to the first-order dynamics model discussed in the main text. The curves have a successive positive shift in the $y$-axis to help visualization. \textbf{(b)} The average fraction of small particles (type 2) in the bottom compartment at steady state, $\langle \mathcal{F}_{bottom}\rangle$, as a function of the angle of the obstacles, $\theta_{obs}$. It is within the range $[0.67,0.70]$, approximately. It features a maximum at $\theta_{obs} = 15^o$. \textbf{(c)} The fraction of open gaps, $\mathcal{P}_{total}$, showing a monotonic increasing with $\theta_{obs}$. Due to the lack of activity, the big particles (type 1) do not get trapped in the funnels, and then the fraction of open gaps is generally large, reaching a value close to $0.90$ for $\theta_{obs} = 75^o$. \textbf{(d)} and \textbf{e} Present the kinetic constants $k^+$ and $k^-$, respectively, as functions of $\theta_{obs}$. Both exhibit a minimum at $\theta_{obs} = 15^o$, and then increase monotonically. $k^+$ is larger than $k^-$ since in the top compartment there is still the effect of steric interactions and area exclusion induced by the big particles. \textbf{(f)} The time constant shows a maximum at $\theta_{obs}$, and then decreases monotonically as a consequence of the increase of the relative fraction of available length for the small particles to flow.\textbf{(g)} The ratio between the diffusion coefficients in the top and bottom compartment along the $y$-direction. The big particles in the top compartment reduce the space of the small particles to diffuse, causing a hindrance of the diffusion coefficient.} 
 \label{fgr:Passive-Passive}
\end{figure*}

\clearpage

\section{Steady state of the active particles vs system's steady state}

\begin{figure*}[h!]
 \centering
 \includegraphics[width=17.0cm]{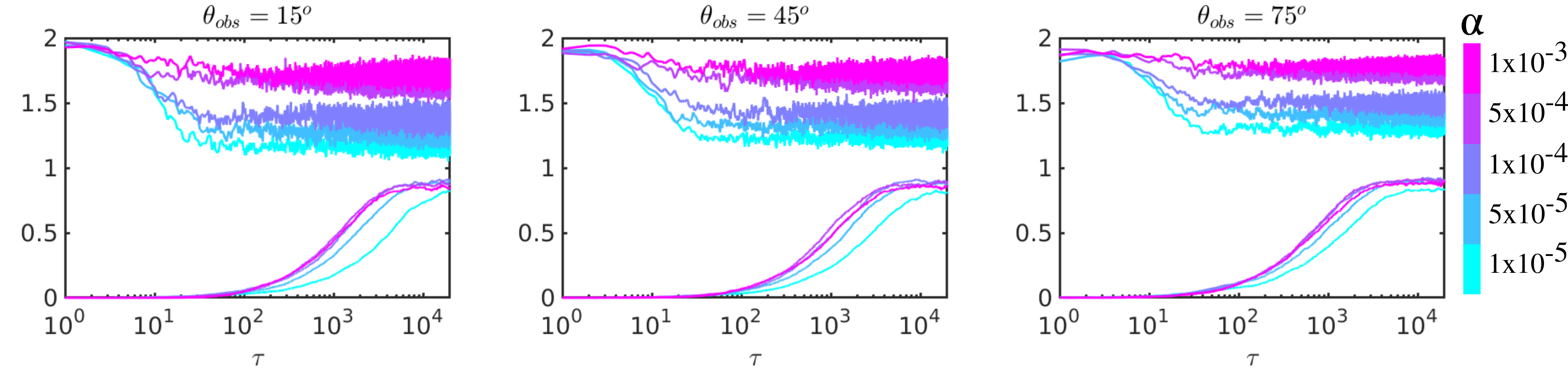}
 \caption{\textit{Active particles time scale vs system's time scale}. As mentioned in the main text, the steady distribution of active particles within the top compartment is reached much faster than the steady state of the whole system. The time response of the whole system is determined by that of $\mathcal{F}_{bottom}$, whereas the time response of the active particles is determined, in a first approximation, by that of $\mathcal{P}_{total}$. The former is presented in the lower part of the figures, whereas the latter is in the top part of the figures due to a positive shift of 1.0 in the $y$-axis to help the visualization. All the parameters of the simulations are those used in the system discussed in the main text. We observe that the time to reach the steady state for $\mathcal{P}_{total}$ is around $\tau = 100$, for all $\alpha$ and $\theta_{obs}$, whereas that of $\mathcal{F}_{bottom}$ is around $\tau = 5000$ for the fastest systems and $\tau = 15000$ for the slowest. The time constant of the system's dynamics is discussed in the main text.}
 
 \label{fgr:SteadyActive_vs_SteadySystem}
 
\end{figure*}

\clearpage

\section{Importance of the asymmetric obstacles }

\begin{figure*}[h!]
 \centering
 \includegraphics[width=\textwidth]{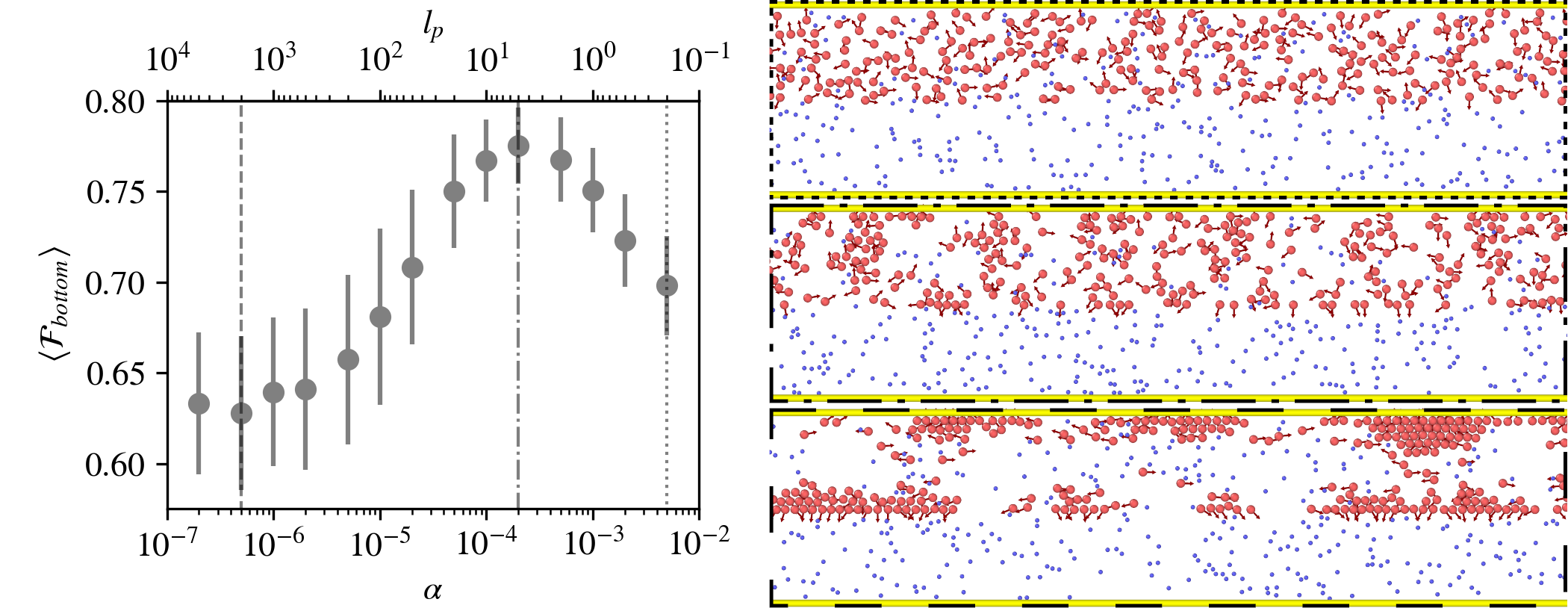}
 \caption{Simulations for the same studied system parameters but considering a flat reflective wall (invisible in the snapshots) only for active particles (red), passive particles (blue) can freely traverse it. The trapping efficiency has a maximum at $\alpha \approx 2\times 10^{-4}$ ($l_p \approx 5 $). When active particles are too persistent ($\alpha \approx 5\times 10^{-7}$, dashed line) they accumulate at walls and are not available for sweeping the top compartment and maximize the number of active kicks given to the passive particles, if they are not enough persistent ($\alpha \approx 5\times 10^{-2}$, dotted line) the number of active kicks also decreases since they reorient more frequently, in between them ($\alpha \approx 2\times 10^{-4}$, dash-dotted line) there is a balance between these two contributions that leads to the observed maximum. This maximum reaches around $100\times \langle \mathcal{F}_{bottom}\rangle \approx 75\%$ of sorting efficiency, so the obstacles are responsible for $\approx 17\%$ of the trapping (compare to fig. 3a in the main text).} 
 \label{fgr:wall_for_active}
\end{figure*}

\clearpage

\section{PDF of displacements}

\begin{figure*}[h!]
 \centering
 \includegraphics[width=15.0cm]{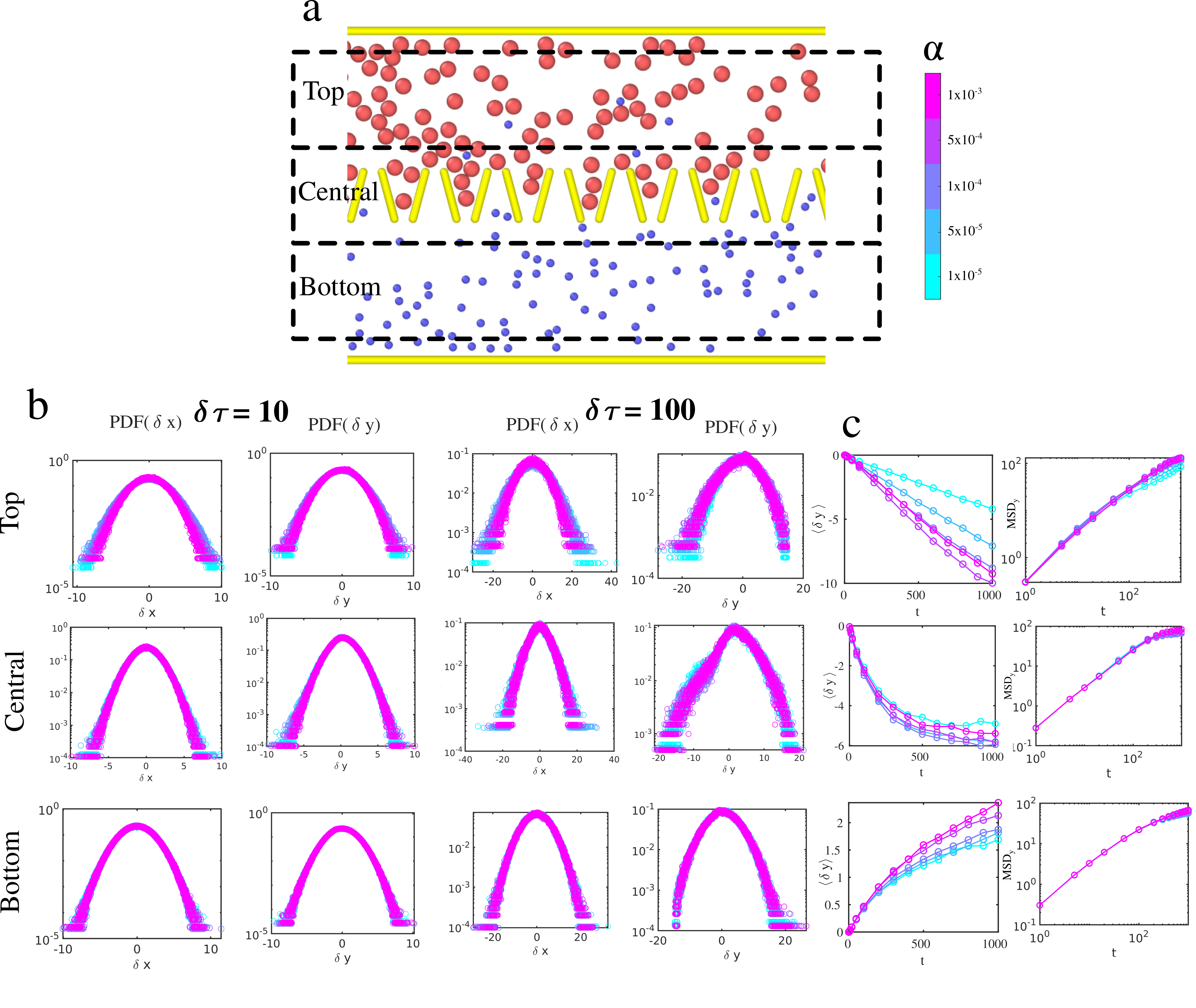}
 \caption{\textit{Probability Density Functions (PDF) of displacements}.\textbf{(a)} The three different regions in which the PDF's were calculated. \textbf{(b)} PDF's in the $x$ and $y$ directions for the different regions and two selected lag times for microchannels with $\theta_{obs} = 45^o$. \textbf{(c)} The mean displacement (drift), $\delta y$ and the mean square displacement (MSD), $\langle\delta y^2\rangle - \langle\delta y\rangle^2$, in the $y$-direction for the three regions considered for $\theta_{obs} = 45^o$. In the PDFs along the $y$ direction, we observe that in the top compartment there is a negative drift induced by the active bath as well as in the central region. The distributions are asymmetric and exhibit shifted tails in the negative $y$-direction; in the bottom compartment there is a drift in the positive direction, induced by the confining wall at $y = 0$. Along the $x$ direction, the PDFs are symmetric in all regions. 
 }

 \label{fgr:PDF}

\end{figure*}

\clearpage

\section{Density profiles}

\begin{figure*}[h]
 \centering
 \includegraphics[width=9.0cm]{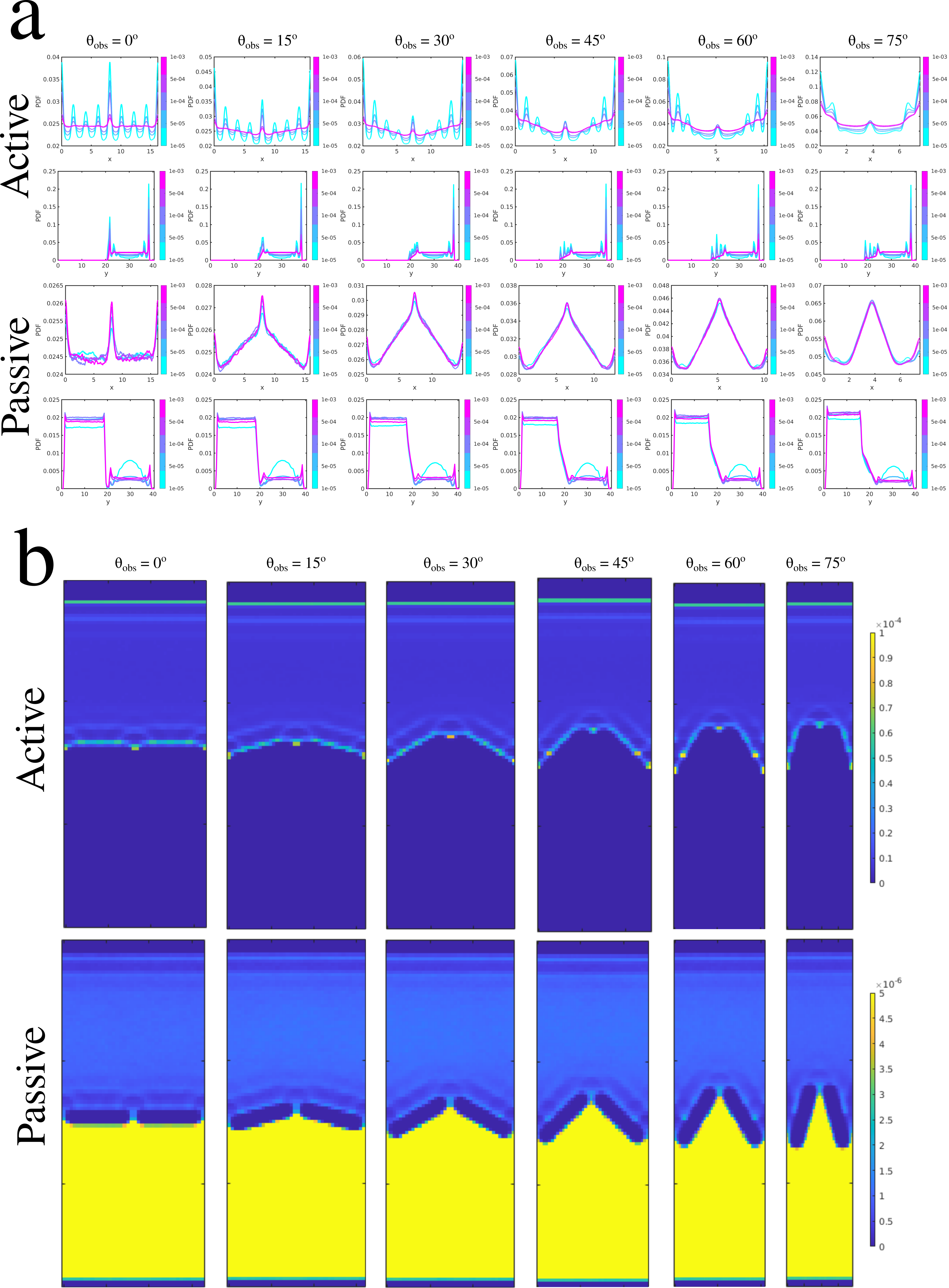}
 \caption{\textit{Normalized Density profiles (PDF)}.\textbf{(a)} One-dimensional PDF in $x$ and $y$ for active and passive particles for all the $\theta_{obs}$ considered and some selected values of $\alpha$ represented in the color bars. \textbf{(a)} The PDF's in the $x$ direction are collapsed to the period of the microchannels given by $\lambda = 2\mathcal{L}\cos(\theta_{obs}) + 2(\Delta + \sigma_{33})$, with the extremes corresponding to a concave gap and the center to a convex gap. \textbf{Active:} the PDF's in the $x$-direction show the shape of a periodic function, reaching maxima at the concave gaps. The height of the peaks decreases as the tumbling rate increases and for high $\theta_{obs}$ the peaks of the PDF's becomes less pronounced and some of them disappear due to the less space available inside the funnels. The PDF's in the $y$-direction shows an accumulation of particles close to the walls and at the funnels. As $\theta_{obs}$ increases, a multilayered structure formed by stacking particles inside the funnels appears. This structure is more pronounced at low $\alpha$. \textbf{Passive:} the PDF's in the $x$-direction reveal that the particles tend to accumulate more close to the convex gaps. This happens because those inside the funnels, and thus close to the concave gaps, are more frequently pushed down by active particles. In the $y$-direction, we observe the same tendency for all $\theta_{obs}$: the PDF's show a small peaks close to the walls and the funnels region in both compartments. In the bulky part of each compartment, the density profile is constant, except for $\alpha \leq 5\times10^{-5}$, case in which the profile in the top compartment shows a rather parabolic shape with a maximum in the center of the compartment. \textbf{(b)} The two-dimensional PDF of active and passive particles collapsed to the period of the microchannel for all the $\theta_{obs}$ considered and for $\alpha = 1\times10^{-4}$. The color map represents the values of the PDF's.}
 
 \label{fgr:PDF}

\end{figure*}

\clearpage

\section{Advection-Diffusion model}
The model introduces an adaptation of that initially proposed in \cite{williams2022confinement}.Alternatively to the main text, we can express the model as:

\begin{equation}
\frac{\partial c(t,y)}{\partial t} = \frac{\partial}{\partial y} \left[ D(y) \frac{\partial c(t,y)}{\partial y} \right] - \frac{\partial}{\partial y} \left[ v(y) c(t,y) \right],
\label{e:1}
\end{equation}

With,

\begin{equation}
D(y) = D_\text{bottom} + \frac{D_\text{top} - D_\text{bottom}}{1 + \exp(-k(y - L_s))}, \label{e:Dsigmoid}
\end{equation}

And introducing the active diffusion term in the drift velocity like,

\begin{equation}
v(y) = \frac{v_{top}}{1 + \exp(-k(y - L_s))} - \frac{dD(y)}{dy}
\label{e:vsigmoid},    
\end{equation}

$L_s = L_y/2$. The corresponding derivatives are:
\begin{align}
\frac{d D}{dy} &= \frac{(D_{\text{top}} - D_{\text{bottom}}) \, k \, \exp\left[-k\left(y - L_s\right)\right]}{\left(1 + \exp\left[-k\left(y - L_s\right)\right]\right)^2} \\
\frac{d v}{d y} &= \frac{v_{\text{top}}  \, k \, \exp\left[-k\left(y - L_s\right)\right]}{\left(1 + \exp\left[-k\left(y - L_s\right)\right]\right)^2} - \frac{d^2 D}{dy^2}
\end{align}

and,

\begin{equation}
\frac{d^2 D}{dy^2} = - (D_{\text{top}} - D_{\text{bottom}}) \, k^2 \, \frac{\exp\left[-k(y - L_s)\right] \left[1 - \exp\left[-2k(y - L_s)\right]\right]}{\left(1 + \exp\left[-k(y - L_s)\right]\right)^4}
\end{equation}

Both formulations are equivalent. We need to take into account that in the passive-passive case, although $D_{top} \neq D_{bottom}$, the gradient in the diffusivity must be set to $0$ since  there is no active diffusion.





\begin{figure*}[h]
 \centering
 \includegraphics[width=12.0cm]{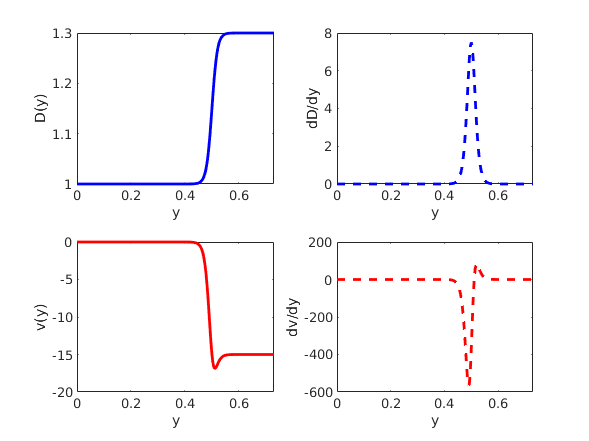}
 \caption{\textit{The space-dependent diffusion coefficient and drift velocity and their derivatives}}
 \label{fgr:derivatives}

\end{figure*}

To solve the PDE, we discretize the spatial domain with a uniform grid: \( y_i = i \, \Delta y \), \( i = 1, \ldots, N \), where \( \Delta y = L/(N-1) \). Let \( c_i^n \approx c(y_i, t^n) \) denote the numerical approximation of the concentration at time \( t^n = n \, \Delta t \).

\begin{equation}
\frac{c_i^{n+1} - c_i^n}{\Delta t} = -\frac{J_{i+1/2}^n - J_{i-1/2}^n}{\Delta y}.
\end{equation}

The fluxes are computed at the cell faces (i.e., midpoints between nodes). The flux at $y_{i\pm1/2}$ is approximated as:

\begin{equation}
J^n_{i\pm 1/2} = -D_{i\pm 1/2} \cdot \frac{c_{i \pm 1}^n - c_i^n}{\Delta y} + v_{i\pm 1/2} \cdot \frac{c_i^n + c_{i \pm 1}^n}{2}
\end{equation}
where $D_{i\pm1/2} = \frac{D_i + D_{i\pm1}}{2}$ and $v_{i\pm1/2} = \frac{v_i + v_{i\pm1}}{2}$.
The boundary conditions are imposed via ghost points, using mirrored values to ensure zero-flux at the boundaries.

\begin{figure*}[h]
 \centering
 \includegraphics[width=9.0cm]{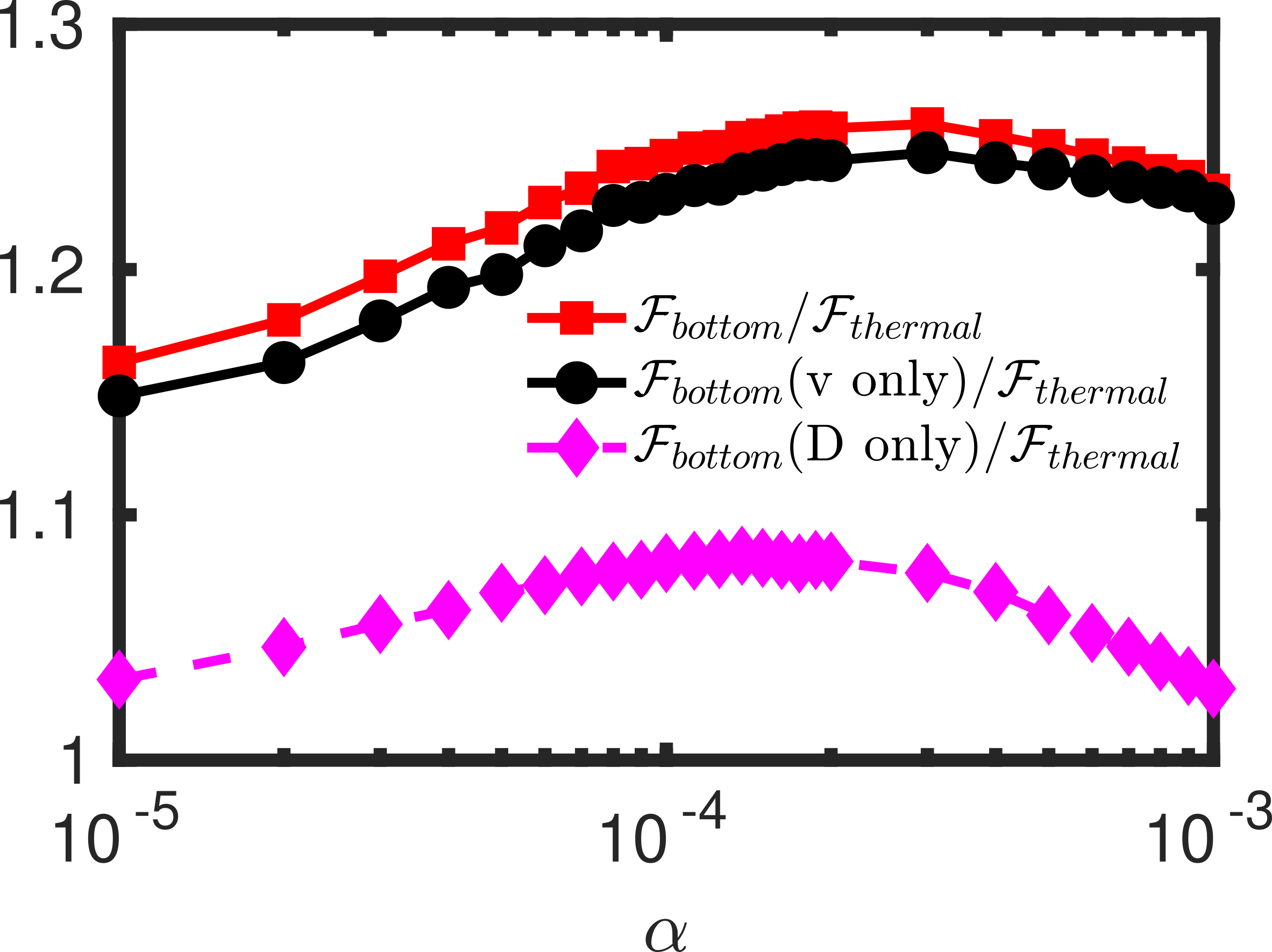}
 \caption{\textit{Drift and diffusivity contributions to $\mathcal{F}_{bottom}$.} Values of $\mathcal{F}_{bottom}$ normalised by its value in the passive-passive case, for the full system (red squares), for thermal diffusivity everywhere but active drift $v_y$ at the top (black circles) and for zero drift everywhere but active diffusivity at the top (magenta diamonds). Values for $\theta_{obs}=75^{\circ}$. The active drift used was the one from the best fit procedure.}
 \label{fgr:F_bottom_vs_F_thermal}

\end{figure*}

\clearpage

\section{Clogging and unclogging times}

\begin{figure*}[htpb!]
 \centering
 \includegraphics[width=14cm]{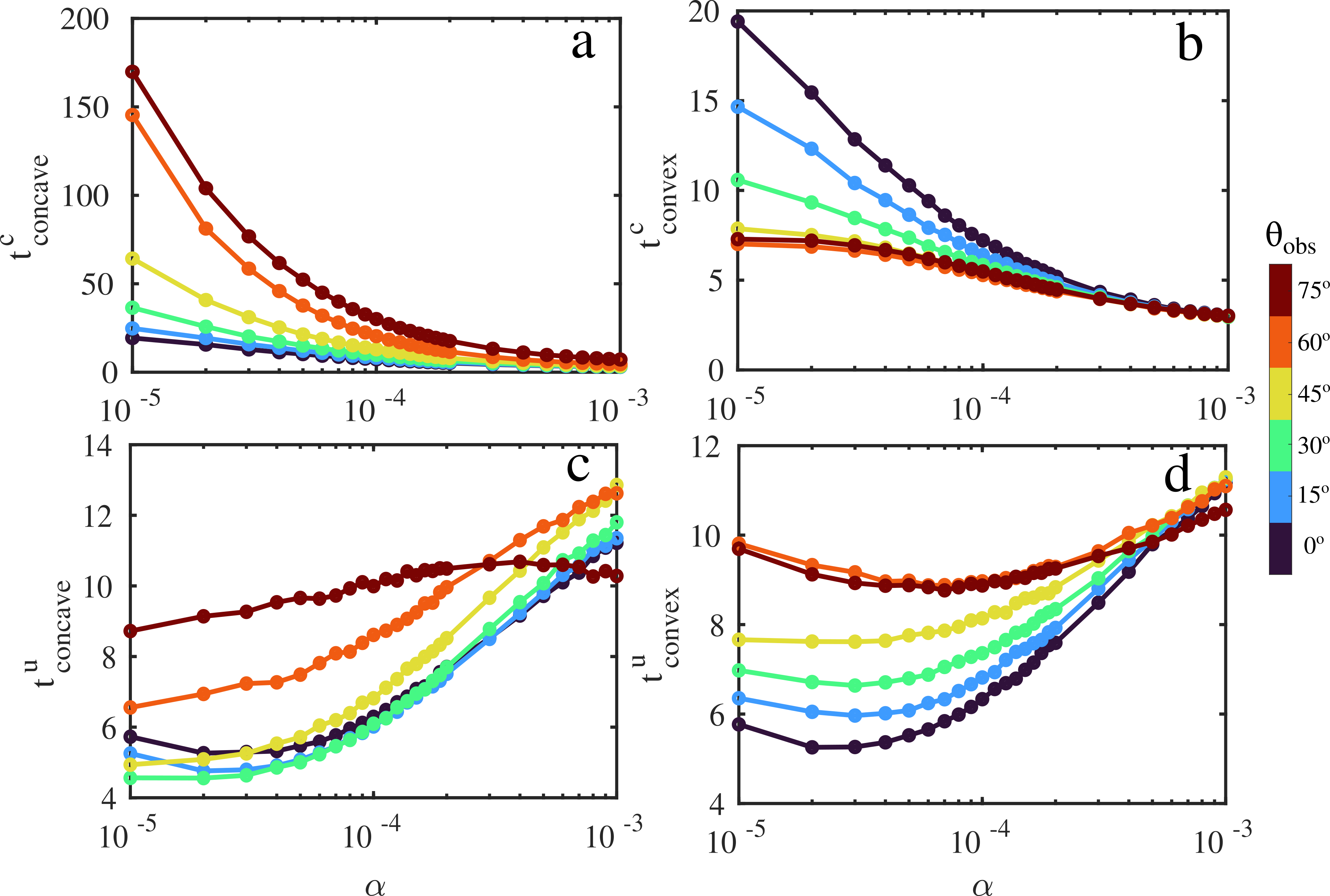}
 \caption{\textit{Clogging and unclogging times for concave and convex gaps}. \textbf{(a)} The average clogging times of concave gaps, $t^c_{concave}$, and \textbf{(b)} convex gaps, $t^c_{convex}$. Both are monotonically decreasing with $\alpha$, a consequence of the increase rate of reorientations leading clogging particles to leave the clogging area, which is reasonable since as $\alpha$ increases the probability of the active particles reorienting in a direction that allows them to leave the gaps also increases and thus the clogging time of the gap decreases. As expected, we observe that $t^c_{concave}$ is larger than $t^c_{convex}$ for all cases, due to the accumulation of active particles at concave regions and thus the subsequent longer time scale for unblocking a concave gap, since it depends depending on the re-orientation of the active particles stacking on top of those one clogging the gap. When it comes to $\theta_{obs}$, however, the behavior of $t^c_{convex}$ and $t^c_{concave}$ is different, being monotonically decreasing for the former and monotonically increasing for the latter. This is much clearer at low tumbling rates ($\alpha \leq 1\times10^{-4}$). At high values of $\alpha$, $t^c_{concave}$ and $t^c_{convex}$ are practically independent of $\theta_{obs}$. The average unclogging times of concave gaps $t^u_{concave}$ \textbf{(c)}, and convex $t^u_{convex}$ \textbf{(d)} show similar time scales for all values of $\alpha$ and $\theta_{obs}$. $t^u_{concave}$ is mainly independent of $\theta_{obs}$ for small values of it($\theta_{obs}<45^{\mycirc}$). Conversely, $t^u_{convex}$ depends strongly on $\theta_{obs}$ for small angles and small values of $\alpha$ whereas it is almost independent of $\theta_{obs}$ for high values of it ($\theta_{obs}\geq60^{\mycirc}$). Thus, we can define an upper and lower limit in $\theta_{obs}$: the lower limit corresponds to $\theta_{obs} = 30^{\mycirc}$ and defines the maximum angle at which $t^u_{concave}$ is independent of the angle, and the upper limit corresponds to $\theta_{obs} = 60^{\mycirc}$ and defines the minimum angle at which $t^u_{convex}$ is independent of the angle. Note that $\mathcal{P}_{concave} = \frac{t^u_{concave}}{t^u_{concave} + t^c_{convex}}$ and $\mathcal{P}_{convex} = \frac{t^u_{convex}}{t^u_{concave} + t^c_{convex}}$  }
 \label{fgr:ResidenceTime}
\end{figure*}

\bibliography{BIOMICAR_SERNA}